\documentclass [aps,12pt,aps,a4paper,prb]{revtex4}
\input epsf
\usepackage{graphicx}
\usepackage{amssymb}
\usepackage{latexsym}

\begin{document}
\begin{center}
\title{\Large\bf Confinement Free Energy of Flexible Polyelectrolytes in Spherical Cavities}
\author{{\bf Rajeev Kumar and M.Muthukumar} \footnote[1]{To whom any correspondence should be addressed, Email : muthu@polysci.umass.edu}}

\affiliation{\it Dept. of Polymer Science \& Engineering, Materials Research Science \& Engineering Center,\\
 University of Massachusetts,
 Amherst, MA-01003, USA.}
\maketitle

\end{center}
\vskip0.1cm
\begin{center}
ABSTRACT
\end{center}
\vskip0.1cm

\noindent A weakly charged flexible polyelectrolyte chain in a neutral spherical cavity is
analyzed using self-consistent field theory (SCFT) within an explicit solvent model.
Assuming the radial symmetry for the system, it is found
that the confinement of the chain leads to creation of a charge density
wave along with the development of a potential difference across the
center of cavity and the surface.
We show that the solvent entropy
plays an important role in the free energy of the
confined system.
For a given radius of the spherical cavity and fixed charge density along the
backbone of the chain,
solvent and small ion entropies dominate over all other contributions
when chain lengths are small. However, with the increase in chain length,
chain conformational entropy and polymer-solvent interaction energy also become important.
Our calculations reveal that energy due to electrostatic interactions
plays a minor role in the free energy. Furthermore, we show that
the total free energy under spherical confinement is not extensive in the
number of monomers. Results for the osmotic pressure and
mean activity coefficient for monovalent salt are presented.
We demonstrate that fluctuations at one-loop level lower the free energy and corrections to the
osmotic pressure and mean activity coefficient
of the salt are discussed. Finite size corrections are shown to widen the
range of validity of the fluctuation analysis.

\clearpage
 \section{INTRODUCTION}
\setcounter {equation} {0}
Polyelectrolytes are ubiquitous in nature and exhibit rich behavior.
In the past, a great deal of theoretical\cite{pincus76,khokhlov80,muthu87,borue88,marko91,muthu96,cruz00, muthu021,muthu02,joanny04} and experimental
efforts\cite{kanaya011,kanaya01,eisenberg96,jenekhe98,prabhu03,forster04} have been made in understanding their characteristics
in solutions with concentrations ranging from dilute to concentrated. One of the remarkable discoveries in the last century is the theory of simple electrolytes made by Debye and H\"{u}ckel\cite{mcquarie}, where electrostatic interactions get screened (colloquially referred to as the Debye screening).
Similar phenomenon was shown to be present in the case of neutral polymers, where monomers interact by short range excluded volume interaction potential (known as Edward's screening\cite{edwardsbook}).
In the case of polyelectrolytes, both kinds of screening effects are present and their coupling via the concept of double screening was introduced by Muthukumar\cite{muthu96}, and the
behavior of polyelectrolyte solutions was described in terms of these
screening phenomena. As a result, different concentration regimes in polyelectrolyte solutions
were predicted and verified experimentally\cite{kanaya011,kanaya01}.

However, most of the computer and real-world experiments
involve finite volume, where boundary effects play a significant
role and can not be ignored. Recently, there has been a resurgence of interest
in studying polymers within confined domains\cite{scalingbook,khokhlovbook,liporna_pnas82,chenmuthu,park96,brandin96,brandin99,muthu03,muthujoey04,murphy07,luijten1,luijten2}. Unlike neutral polymers, little is
known about the physics of polyelectrolytes under confinement, a situation realized in liposome-mediated delivery of macromolecules to the cells\cite{liporna_pnas82},
translocation experiments involving RNA/DNA\cite{brandin96,brandin99}, synthetic polyelectrolytes\cite{murphy07}
  etc. Underlying physics in these experiments is governed by the confinement effects on
a \textit{single} polyelectrolyte chain. Physically, confinement forces interaction among
the monomers and the conformational entropy of the chain gets lowered due to less
number of conformational states available to the chain. If small components like
salt ions and solvent molecules are also present in addition to the polymer,
then translational entropy of these components
gets diminished, making confinement a thermodynamically unfavorable
process. Quantitative estimates of these confinement effects are desirable.
Recently, a quantitative description of the
finite size corrections\cite{finite_tellez} to
free energy for electrolytic systems 
has been presented. The analog
of these calculations for polyelectrolyte systems has not been attempted yet and is one of the
goals of this study.

In this study, we focus on a single polyelectrolyte chain confined in a neutral spherical cavity and use
radial symmetry to obtain the mean field results. This theoretical model is pertinent in understanding many
important physical processes. Few promising applications of this model are the
computation of free energy barriers for the chain to move out of the confining space,
osmotic pressure of polyelectrolytes, etc. To start with, we consider a situation where
inner and outer dielectric constants of medium are different (say $\epsilon_{i}$ and $\epsilon_{o}$, respectively). For this situation, the electrostatic potential is to be calculated using continuity of electrostatic potential and normal component of displacement vector at the boundary. It turns out that these boundary conditions are equivalent  to a continuous dielectric medium (of dielectric constant $\epsilon_{i}$ everywhere) with charges confined within the boundary of sphere for the {\it radially symmetric globally electroneutral} system. In other words, dielectric mismatch effects disappear due to the use of radial symmetry. We must point out that radial symmetry is strictly
valid only at the mean field level and fluctuations break this symmetry. In the concentrated regime, we have been
able to compute the fluctuation contributions without recourse to radial symmetry.

Unlike cylindrical\cite{scalingbook} and
rectangular\cite{scalingbook,chenmuthu} confinements, a single self-avoiding chain in a spherical cavity
has been shown to be a polymer solution problem\cite{edwardsbook,luijten1,luijten2,khokhlovbook,muthuedwards,janninkbook}
with different degrees of confinements corresponding to different concentration regimes as seen in
polymer solution theories. We are here interested in estimating different thermodynamic contributions to
the free energy of confined polyelectrolytic system. Computing these
contributions using simulations is a formidable task. However,
self-consistent field theory (SCFT)
presents a faster and an accurate way to address this problem. SCFT allows us to explore the
role of confinement in free energy at the mean field level (also known as the saddle point approximation)
and then capture the role of fluctuations by expanding free energy functional
around the mean field solution (one-loop calculations).
Unlike earlier studies on single neutral chain\cite{muthujoey04}, we use
``explicit solvent model", which captures solvent entropy in a
more realistic way. At the mean field level, free
 energy for a self-avoiding chain under strong confinement (in vacuum)
 has been shown to be proportional to  $w \bar{\rho}_{p}^{2} \Omega$, where
 $w$ is a measure of the strength of excluded volume interactions, $\bar{\rho}_{p}$ is
 monomer density and $\Omega$ is the volume of the cavity. Remarkably, same proportionality
 is exhibited by polymer solution theories\cite{edwardsbook,muthuedwards} in the concentrated
regime. However, at {\it low} polymer concentrations (i.e. dilute and semi-dilute regime), fluctuations
around the saddle point solution become important and saddle point approximation breaks down.
In that case, non-perturbative techniques have to be devised to compute the correct free energy.
In this work, we explore the free energy in the concentrated regime, when radius of
gyration of the chain is comparable to the radius of the cavity and mean field theory is still
applicable.

Dividing the free energy into energetic and entropic parts using thermodynamic arguments, we have
identified the role of individual components. Moreover, one loop calculations provide insight about
the corrections to the bulk expressions from fluctuation effects in the concentrated regime.
For a globally neutral system with only small ions
(i.e. without polyelectrolyte) inside a neutral cavity, local electroneutrality is the equilibrium state.
In contrast, local electroneutrality is violated in the presence of a polyelectrolyte chain
due to the depletion effects present in the system. We have studied
the resulting monomer and charge density distribution for different sets of relevant
parameters of the problem. To link with the experiments, we have computed the osmotic
pressure and mean activity coefficient for monovalent salt.

This paper is organized as follows: theory is presented in Sec. ~\ref{sec:theory}; calculated results and conclusions are presented in Sec. ~\ref{sec:results} and ~\ref{sec:conclusions}, respectively.

\section{Theory}
\setcounter {equation} {0} \label{sec:theory}
\subsection{Self-consistent field theory} \label{sec:theory_scf}

We consider a spherical cavity of radius $R$ containing
a single flexible polyelectrolyte chain of
total $N$ Kuhn segments, each with length $b$.
The polyelectrolyte chain is represented
as a continuous curve of length $Nb$ and an arc length variable $t$
is used to represent any segment along the backbone. We assume that there are $n_{c}$
monovalent counterions (positively charged) released by the chain (and assuming that the chain is negatively charged for the sake of specificity).
In addition, there are $n_{\gamma}$ ions
of species $\gamma \,(= +,-)$ coming from added salt (in
volume $\Omega = 4 \pi R^{3}/3$ ) so that the whole system is globally electroneutral. Let $Z_{j}$ be
the valency (with sign) of the charged species of type $j$. 
Moreover, we assume that there are $n_{s}$ solvent molecules (satisfying
the incompressibility constraint) present
in the cavity and for simplicity, each solvent molecule occupies a volume ($v_{s}$)
same as that of the monomer (i.e. $v_{s}\equiv b^{3}$).
Subscripts $p,s,c,+$ and $-$ are used to represent
monomer, solvent , counterion from  polyelectrolyte, positive and negative salt ions, respectively. Degree of ionization of the chain is taken to be $\alpha$ and we consider smeared charge
distribution so that each of the segments carries a charge $e\alpha Z_{p}$, where $e$
is the electronic charge. 

If electrostatic interactions between solvent molecules and
small ions are ignored, the partition function for the system can be
written as
\begin{eqnarray}
       \mbox{exp}\left(-\frac{F} {k_{B}T}\right )& = & \frac {1}{\prod_{j}n_{j}!}\int D[\mathbf{R}] \int \prod_{j} \prod_{m=1}^{n_{j}} d\mathbf{r}_{m} \quad \mbox{exp} \left \{-\frac {3}{2 b}\int_{0}^{Nb}
        dt\left(\frac{\partial \mathbf{R}(t)}{\partial t} \right )^{2} \right . \nonumber \\
&& -  \frac {1}{2 b^{2}} \int_{0}^{Nb}dt \int_{0}^{Nb}dt' V_{pp} [ \mathbf{R}(t) - \mathbf{R}(t')]
 -\frac{1}{b}\sum_{j}\sum_{m =1 }^{n_{j}}  \int_{0}^{Nb}dt V_{pj} [ \mathbf{R}(t) - \mathbf{r}_{m}]
\nonumber \\
&& \left .      - \frac{1}{2}\sum_{j}\sum_{a}\sum_{m =1 }^{n_{j} }\sum_{p =1 }^{n_{a}} V_{ja} [ \mathbf{r}_{m} - \mathbf{r}_{p}] \right \}\prod_{\mathbf{r}}\delta\left(\hat{\rho}_{p}(\mathbf{r}) + \hat{\rho}_{s}(\mathbf{r}) - \rho_{0}
\right) \label{eq:parti_sing}
\end{eqnarray}
where $\mathbf{R}(t)$ represents the position vector for $t^{th}$ segment and subscripts $j,a = s,c,+,-$. In the above expression, it is understood that the factor of $1/2$ in the last term
inside the exponent is there, only when $j=a$.
In Eq. (~\ref{eq:parti_sing}), $k_{B}T$ is the Boltzmann constant times absolute
temperature. $V_{pp}(\mathbf{r}),V_{ss}(\mathbf{r}) \mbox{ and}\: V_{ps}(\mathbf{r})$ represent
monomer-monomer, solvent-solvent and monomer-solvent interaction energies, respectively,
when the interacting species are separated
by distance $r = \mid \mathbf{r} \mid $ and are given by
\begin{eqnarray}
       V_{pp}(\mathbf{r}) & = & w_{pp} \delta(\mathbf{r}) + \frac {Z_{p}^{2}e^{2}\alpha^{2}}{\epsilon
       k_{B}T}\frac{1}{r} \\
       V_{ss}(\mathbf{r}) & = & w_{ss} \delta(\mathbf{r}) \\
       V_{ps}(\mathbf{r}) & = & w_{ps} \delta(\mathbf{r}).
\end{eqnarray}

Here, $w_{pp}, w_{ss}$ and $w_{ps}$ are the excluded volume parameters, which characterize
the short range excluded volume interactions between monomer-monomer, solvent-solvent and monomer-solvent,
respectively. $\delta(\mathbf{r})$ is the three dimensional
Dirac delta function and $\epsilon$ is position independent effective dielectric constant
of the medium ( in units of $ 4 \pi \epsilon_{o}$, where $\epsilon_{o}$ is
the permittivity of vacuum). Delta functions involving microscopic densities
enforce incompressibility condition at all points in the system ($\rho_{0}$
being total number density of the system). Interactions between polyelectrolyte segments and small ions as
represented by $V_{pj}$, are given by
       \begin{eqnarray}
       V_{pj}(\mathbf{r}) & = & \frac {Z_{p}Z_{j}e^{2}\alpha }{\epsilon k_{B}T}\frac{1}{r}\quad \mbox{for} \quad j = c,+,-.
       \end{eqnarray}

 In writing the interaction energies between polyelectrolyte segments and small ions,
we have taken small ions to be point charges so that
they have zero excluded volume and interactions are purely electrostatic in nature.
In the point charge limit for the small ions, interaction between two small ions is
written as

       \begin{eqnarray}
       V_{ja}(\mathbf{r}) & = & \frac {Z_{j}Z_{a}e^{2}}{\epsilon k_{B}T}\frac{1}{r} \quad \mbox{for} \quad j,a = c,+,-.
\end{eqnarray}

Following the standard protocol\cite{freedbook} to obtain saddle point equations (see Appendix A), the Poisson-Boltzmann equation for electric potential gets coupled to the well-known modified diffusion equation for chain connectivity. In particular, the saddle point equations are given by

\begin{eqnarray}
\phi_{p}(\mathbf{r}) &=& \chi_{ps}b^{3}\rho_{s}(\mathbf{r}) + \eta(\mathbf{r})\label{eq:saddle_exp1}\\
\phi_{s}(\mathbf{r}) &=& \chi_{ps}b^{3}\rho_{p}(\mathbf{r}) + \eta(\mathbf{r})\\
\rho_{p}(\mathbf{r})&=&  \frac{1}{Q_{p}}\int_{0}^{N} ds \, q(\mathbf{r},s) q(\mathbf{r},N-s) \label{eq:den_eq}\\
\rho_{s}(\mathbf{r})&=& \frac{n_{s}}{Q_{s}}\mbox{exp}\left [-\phi_{s}(\mathbf{r}) \right]\\
\rho_{p}(\mathbf{r}) &+& \rho_{s}(\mathbf{r}) = \rho_{0} \\
\rho_{j}(\mathbf{r})&=& \frac{n_{j}}{Q_{j}}\mbox{exp}\left [- Z_{j}\psi(\mathbf{r}) \right] \quad \mbox{for}\quad j = c,+,-\label{eq:den_small}\\
\bigtriangledown_{\mathbf{r}}^{2}\psi(\mathbf{r}) &=& - 4 \pi l_{B} \left [ \sum_{j = c,+,-} Z_{j}\rho_{j}(\mathbf{r}) + Z_{p}\alpha \rho_{p}(\mathbf{r})\right ]. \label{eq:saddle_last}
\end{eqnarray}

These equations are equivalent to those derived by Shi and Noolandi\cite{shi}, and Wang\cite{wang} \textit{et al.}, although the method of derivation is different as briefly outlined in Appendix A. In these equations, $\rho_{\beta}(\mathbf{r})$ and $\phi_{\beta}(\mathbf{r})$ are respectively the macroscopic number density and the field experienced by particles of type $\beta$, due to excluded volume interactions. All charged species experience electrostatic potential
represented by $\psi(\mathbf{r})$ above.   Note that $\psi(\mathbf{r})$ in above equations is dimensionless
(in units of $k_{B}T/e$) and $l_{B}$ depicts the Bjerrum length defined as $l_{B} = e^{2}/\epsilon k_{B}T$. Moreover, $\chi_{ps}$ is
the dimensionless Flory's chi parameter and $\rho_{0} = (N + n_{s})/\Omega = 1/b^{3}$.
Also, $\eta(\mathbf{r})$ is the
well-known pressure field introduced to enforce the incompressibility
constraint.
The function $q(\mathbf{r},s)$ is the probability of finding segment $s$ at location $\mathbf{r}$, when starting end of the chain can be anywhere in space, satisfying the modified diffusion equation\cite{helfand75}
\begin{eqnarray}
\frac{\partial q(\mathbf{r},s) }{\partial s} &=& \left [\frac{b^{2}}{6}\bigtriangledown_{\mathbf{r}}^{2}- \left \{ Z_{p}\alpha \psi(\mathbf{r}) + \phi_{p}(\mathbf{r})\right \} \right ]q(\mathbf{r},s), \quad s\in (0,N). \label{eq:difreal}
\end{eqnarray}
Also, $Q_{\beta}$ represents the partition function for the particle of type $\beta$ in the field experienced by it, given by

\begin{eqnarray}
Q_{s} &=& \int d\mathbf{r} \, \mbox{exp}\left [ -\phi_{s}(\mathbf{r}) \right ] \label{eq:saddle_explast}\\
Q_{p} &=& \int d\mathbf{r} q(\mathbf{r},N) \label{eq:pe_green}\\
Q_{j} &=& \int d\mathbf{r} \, \mbox{exp}\left [ - Z_{j}\psi(\mathbf{r}) \right ]\quad \mbox {for}\quad j = c,+,-.\label{eq:small_green}
\end{eqnarray}
Using the above equations, approximated free energy at the extremum (saddle point approximation) is given by
\begin{eqnarray}
\frac{F^{\star}}{k_{B}T} &=& \frac{F_{0}}{k_{B}T} - \chi_{ps}b^{3}\int d\mathbf{r}  \rho_{p}(\mathbf{r})\rho_{s}(\mathbf{r}) + \frac{1}{8\pi l_{B}}\int d\mathbf{r} \psi(\mathbf{r})\bigtriangledown_{\mathbf{r}}^{2}\psi(\mathbf{r}) - \mbox{ln}Q_{p} +
\sum_{j}n_{j}\left [ \mbox{ln}\frac{n_{j}}{Q_{j}} - 1\right ] \nonumber
\\&& - \rho_{0}\int d\mathbf{r} \eta(\mathbf{r}), \label{eq:free_energy}
\end{eqnarray}
where $j=s,c,+,-$ and $F_{0}/k_{B}T = \frac{\rho_{0}}{2} \left( Nw_{pp} + n_{s}w_{ss} \right )$ is the self-energy
contribution arising from excluded volume interactions.  
 Using thermodynamic arguments\cite{marcus55} and assuming dielectric constant ($\epsilon$) to be independent of temperature ($T$), the free energy (Eq. (~\ref{eq:free_energy})) is divided into
enthalpic contributions due to excluded volume, electrostatic interactions and
entropic part due to small ions, solvent molecules and the polyelectrolyte chain.
Denoting these contributions by $E_{w},E_{e}, S_{ions}, S_{solvent}$ and  $S_{poly}$, respectively, the free energy is written as
\begin{eqnarray}
\frac{F^{\star} - F_{0}}{k_{B}T} &=& \frac{E_{w}}{k_{B}T} + \frac{E_{e}}{k_{B}T} - \frac{T (S_{ions} + S_{solvent} + S_{poly})}{k_{B}T}  \label{eq:free_thermo_exp}\\
\frac{E_{w}}{k_{B}T} &=& \chi_{ps}b^{3}\int d\mathbf{r}  \rho_{p}(\mathbf{r})\rho_{s}(\mathbf{r})  + \rho_{0}\int d\mathbf{r} \eta(\mathbf{r}) \label{eq:ew_exp}\\
\frac{E_{e}}{k_{B}T} &=& \frac{1}{2} \int d\mathbf{r} \,\psi(\mathbf{r})\left (\sum_{j = c,+,-}Z_{j}\rho_{j}(\mathbf{r}) + Z_{p}\alpha \rho_{p}(\mathbf{r})\right ) \label{eq:ee}\\
- \frac{T S_{ions}}{k_{B}T} &=&  - \sum_{j = c,+,-} \left [n_{j} \mbox{ln} Q_{j} +   \int d\mathbf{r}\, Z_{j}\rho_{j}(\mathbf{r}) \psi(\mathbf{r})   \right ]  + \sum_{j = c,+,-}n_{j}\left [ \mbox{ln} n_{j} - 1\right ]\nonumber\\
&=& \sum_{j=c,+,-}\int d\mathbf{r}\,\rho_{j}(\mathbf{r})\left\{\ln\left[\rho_{j}(\mathbf{r})\right] - 1\right \}\label{eq:tsmall}\\
-\frac{T S_{solvent}}{k_{B}T} &=&  - \left [n_{s} \mbox{ln} Q_{s} +  \int d\mathbf{r}\, \rho_{s}(\mathbf{r}) \phi_{s}(\mathbf{r})  \right ]  + n_{s}\left [ \mbox{ln} n_{s} - 1\right ]\nonumber\\
&=& \int d\mathbf{r}\,\rho_{s}(\mathbf{r})\left \{ \ln\left[\rho_{s}(\mathbf{r})\right]- 1 \right \}\label{eq:solvent_ent} \\
- \frac{T S_{poly}}{k_{B}T} &=&  - \ln Q_{p} - \int d\mathbf{r}\, \left [ \left\{Z_{p}\alpha \psi(\mathbf{r}) + \phi_{p}(\mathbf{r}) \right \} \rho_{p}(\mathbf{r}) + \rho_{0}\eta(\mathbf{r})\right ]. \label{eq:poly_ent}
\end{eqnarray}

So far, we have presented a general field theoretical treatment for a single polyelectrolyte chain and haven't considered
confinement. We study the role of confinement by solving Eqs. (~\ref{eq:saddle_exp1}-~\ref{eq:saddle_last}) under
a particular set of boundary conditions and constraints, which are presented in the next section. Also, the limits of volume integral in Eqs. (~\ref{eq:free_thermo_exp}-~\ref{eq:poly_ent}) vary
over the volume of confining spherical cavity.

\subsection{Boundary Conditions and Constraints}
The above treatment leads to coupling of non-linear Poisson-Boltzmann equation with modified diffusion equation. Both of
these equations are second order differential equations and hence, two conditions are required for each, in order to
obtain a unique solution. In addition, for $s$ dependent diffusion equation, an initial condition is needed to start
the computations. To solve these equations, we exploit the assumed spherical symmetry of the system so that $q(\mathbf{r},s) \rightarrow q(r,s)$ and the Laplacian is given by
\begin{eqnarray}
\bigtriangledown_{\mathbf{r}}^{2} &=& \frac{1}{r^{2}}\frac{\partial}{\partial r}\left(r^{2}\frac{\partial}{\partial r}\right)  \label{eq:radlap}
\end{eqnarray}
Due to symmetry of the system, additional requirements need to be fulfilled by
the solution. Here, we summarize all these conditions:

\begin{eqnarray}
       &\mbox{Boundary Conditions :}& \frac{\partial \psi(r)}{\partial r}\mid_{r = R} = 0, q(R,s) = 0 \quad \mbox{for all $s$} \\
       &\mbox{Initial Conditions :} &  q(r,0) = 1 \quad \mbox{for all $r\neq R$} \\
       &\mbox{Symmetry Conditions:} &  \frac{\partial \psi(r)}{\partial r}\mid_{r=0} = \frac{\partial q(r,s)}{\partial r}\mid_{r=0} = 0    \quad \mbox{for all $s$}
\end{eqnarray}
Boundary and initial conditions for $q(r,s)$ correspond to the facts
that (a) segments are excluded from boundary so
probability of finding any segment at the boundary is zero and (b) the ends can be
anywhere inside sphere.
Symmetry condition for $q(r,s)$
is invoked because we are looking for a symmetrical solution
of monomer density about the center without any discontinuity.
Boundary condition and symmetry conditions for electrostatic potential are obtained by
using the Gauss law at the boundary and the fact that net force experienced by an
ion at the center of the sphere must be zero.

Along with the above initial and boundary conditions, solution of SCF equations
need to be obtained
under additional constraints due to the fixed number of monomers ($N$), ions ($n_{j}$) and global electroneutrality
so that
\begin{equation}
Z_{p}\alpha N + Z_{c} n_{c} = 0, \quad Z_{+} n_{+} = - Z_{-} n_{-} = 0.6023 \, c_{s} \Omega,
\end{equation}
where $c_{s}$ is salt concentration in moles per liter (molarity) and $\Omega$ is in units of $\mbox{nm}^{3}$.

\subsection{Numerical Technique}
We have solved SCF equations (Eqs. (~\ref{eq:saddle_exp1} - ~\ref{eq:difreal})) in real space using an
explicit finite difference scheme
for Poisson-Boltzmann-like equation (Eq. (~\ref{eq:saddle_last})) and the standard Crank-Nicolson\cite{recipebook} scheme for solving modified diffusion equation (Eq. (~\ref{eq:difreal})). Due to the use of properly normalized equations for densities, all the constraints mentioned earlier are always satisfied during the computation.

As the solution of SCF equations is invariant when an arbitrary constant is added to
the fields, so this constant needs to be fixed in order to obtain a unique solution.
Choice of fixing this constant depends on the numerical scheme used in
solving the SCF equations. We simply choose $\psi(R) = 0$
and $\int d\mathbf{r} \eta(\mathbf{r}) = 0$. We must point out that
any method of fixing this constant does not affect the densities
but the free energy gets changed by a constant.

To realize the constraint $\int d\mathbf{r} \eta(\mathbf{r}) = 0$,
$\frac{1}{\Omega}\int d\mathbf{r} \eta(\mathbf{r})$ is subtracted from the computed
$\eta(\mathbf{r})$ at each iteration. This procedure leads to $\int d\mathbf{r} \eta(\mathbf{r}) = 0$
in the final solution. Also, all the integrals are evaluated by the standard
Trapezoidal\cite{recipebook} rule
and Broyden's method\cite{recipebook} is used to solve the set of non-linear equations.

\subsection{Reference System} \label{sec:reference_sys}
The choice of a reference system in free energy calculations depends on
the physical quantity of interest.
One of our goals in this study is to investigate the
role of the polyelectrolyte chain in
the free energy of the system. To study the role of the chain, the spherical
cavity without any polymer (with small ions and solvent inside) is the
appropriate choice.

In the absence of the chain, free energy becomes
\begin{eqnarray}
\frac{F\{\bar{\rho}_{p} = 0\} - F_{0}\{N=0\}}{k_{B}T} &=& \sum_{j=+,-,s}n_{j}\left[\ln \frac{n_{j}}{\Omega} - 1\right].
\end{eqnarray}
Note that in the absence of the chain, $n_{s}b^{3} = \Omega$, due to the incompressibility condition. 

 \subsection{Osmotic Pressure: Contact Value Theorem }
 Although we have computed the free energy of a confined chain, it is worthwhile to compute a physical observable,
 which is readily measurable experimentally. So, we have computed the osmotic pressure
 of a confined chain by carrying out the variation of free energy (Eq. (~\ref{eq:free_energy})) with respect to
 the number of solvent molecules, but keeping the number of monomers and salt ions fixed ( and taking care of the fact that volume has to be changed to alter the number of solvent molecules for the incompressible system under investigation here). The osmotic pressure is given by\cite{janninkbook}

\begin{eqnarray}
\frac{\Pi b^{3}}{k_{B}T}&=&  - \left[\frac{\delta }{\delta n_{s}}\left(\frac{F - F_{0}}{k_{B}T}\right)_{N,n_{j}} - \frac{\delta }{\delta n_{s}}\left(\frac{F - F_{0}}{k_{B}T}\right)_{N,n_{j} = 0}\right ],
\end{eqnarray}
where $j=c,+,-$. Within the saddle point approximation, $F = F^{\star}$ and $(F^{\star} - F_{0})/k_{B}T = -n_{s}$ for pure solvent (i.e. when $N,n_{j} = 0 $). Using the above formula,
osmotic pressure comes out to be (within radial symmetry)
\begin{eqnarray}
\frac{\Pi^{\star}}{k_{B}T}&=&  \sum_{j=c,+,-}\rho_{j}(R^{-}) - \rho_{p}(R^{-}) - \ln \rho_{s}(R^{-}) + \frac{q(R^{-},N)}{\int d\mathbf{r}\, q(\mathbf{r},N)} \nonumber \\
&& - \chi_{ps}b^{3}\rho_{p}^{2}(R^{-}) + \rho_{e}(R^{-})\psi(R^{-}), \label{eq:osmotic}
\end{eqnarray}
where $\rho_{e}(R^{-})$ is the total charge density (in units of electronic charge) at a point close to the surface of the cavity and $\star$ in the superscript depicts the fact that saddle point approximation for the free energy has been used in deriving the result. Due to the
coarse grained model used in studying the single chain, information at a length scale below Kuhn's segment
length is not correctly captured by the model. So, $R^{-}$ represents the point, which is at a distance of one Kuhn segment length from the surface. This point is well-discussed in the literature\cite{podgornik_contact,muthu_ho} and will not be pursued further.

If we were to imagine a system, where densities and fields are constant (as in the concentrated bulk system), then the above expression simplifies to the well-known\cite{edwardsbook} expression for the osmotic pressure of a homogeneous system (represented by $\Pi^{\star}_{h}$)
\begin{eqnarray}
\frac{\Pi^{\star}_{h} }{k_{B}T}&=&  \sum_{j=c,+,-}\frac{n_{j}}{\Omega} - \frac{\bar{\rho}_{p}}{b^{3}} - \ln \frac{1-\bar{\rho}_{p}}{b^{3}} + \frac{\bar{\rho}_{p}}{Nb^{3}}- \frac{\chi_{ps}}{b^{3}}\bar{\rho}_{p}^{2}, \label{eq:osmotic_homo}
\end{eqnarray}
where $\bar{\rho}_{p} = Nb^{3}/\Omega$. In the literature\cite{podgornik_contact}, osmotic pressure for fluids near surfaces is given by the
 fluid densities near surfaces (the so called ``contact value theorem'') and Eq. (~\ref{eq:osmotic}) is nothing but
 the analog of the ``contact value theorem'' for the inhomogeneous system \textit{with interactions}.

 \subsection{Mean Activity Coefficient }
 Using the saddle point approximation for the free energy, we have computed the electrostatic chemical potential
 of small ions. Carrying out the variation of free energy (Eq. (~\ref{eq:free_energy})) with respect to number of small ions, the mean field estimate for the electrostatic chemical potential\cite{mcquarie} of small ions comes out to be
\begin{eqnarray}
\mu_{j}^{el}&=& \frac{\delta }{\delta n_{j}}\left(\frac{F - F_{0}}{k_{B}T}\right)_{N,V,n_{m\neq j}} = \ln \left[\frac{n_{j}}{\int d\mathbf{r}\, \mbox{exp}\,\left(-Z_{j}\psi(\mathbf{r})\right)}\right] \quad \mbox{for}\quad j = c,+,-.
\end{eqnarray}

 This is a straightforward generalization of the electrostatic chemical potential for homogeneous system to an inhomogeneous one. Using these expressions, chemical potential for the salt $A_{\nu_{+}}B_{\nu_{-}}$ (so that $\nu_{+} = \nu_{-} = 1$ for the monovalent salt) can be written as
\begin{eqnarray}
\mu_{salt} &=& \nu_{+}\mu_{+}^{el} + \nu_{-}\mu_{-}^{el}. \label{eq:chem_salt}
\end{eqnarray}
 However, individual activity coefficients or the chemical potential can not be measured experimentally. So,
 we construct the mean activity coefficient ($\gamma_{\pm}$) for the binary salt\cite{mcquarie} defined by
\begin{eqnarray}
\mu_{salt} &=& \ln \left[\left(\frac{n_{+}}{\Omega}\right)^{\nu_{+}} \left(\frac{n_{-}}{\Omega}\right)^{\nu_{-}}\gamma_{\pm}^{\nu}\right], \label{eq:mean_salt}
\end{eqnarray}
where $\nu = \nu_{+} + \nu_{-}$. Using Eq. (~\ref{eq:chem_salt}) and  (~\ref{eq:mean_salt}), mean activity coefficient is given by
\begin{eqnarray}
\gamma_{\pm}^{\nu} &=& \left[\frac{\Omega}{\int d\mathbf{r}\, \mbox{exp}\,\left(-Z_{+}\psi(\mathbf{r})\right)}\right ]^{\nu_{+}}\left[\frac{\Omega}{\int d\mathbf{r}\, \mbox{exp}\,\left(-Z_{-}\psi(\mathbf{r})\right)}\right ]^{\nu_{-}}. \label{eq:mean_activity}
\end{eqnarray}
Using Schwarz's inequality\cite{seriesbook}, it can be shown that the mean activity coefficient is always less
than or equal to unity.
For the spherical cavity with pure solvent and monovalent salt ions,
potential is essentially constant everywhere (local electroneutrality) and that leads to mean
activity coefficient being unity. Note that
the Eq. (~\ref{eq:mean_activity}) can also be derived by considering a Donnan equilibrium between the interior containing polyelectrolyte chain and the exterior containing salt ions with salt concentration $c_{s}$.
 \section{Results}
\setcounter {equation} {0} \label{sec:results}
Having presented the field theoretical treatment of a single flexible chain in the presence of solvent, we present the results obtained after solving SCFT equations for a
negatively charged chain ($Z_{p} = -1$) with monovalent salt. In this
study, we have taken all small ions (counterions and co-ions) to be point charges and while
calculating counterion density profiles, we have added contributions coming from the counterions released by
the polyelectrolyte chain and the salt. Also, it should be
noticed that due to the point nature of these charges, counterions and co-ions are not excluded from
the confining boundary, which can be done, in principle, by the introduction of an arbitrary wall potential\cite{duncan99}.
Here, we simply assume that the {\it depletion layer}\cite{fleerbook} for counterions and co-ions is very small as compared to
the monomer due to an order of difference in their sizes and its negligible effects on the system properties.

All the results reported in this paper have been obtained with a grid spacing of $\Delta r = 0.1$ and chain contour discretization of $\Delta s = 0.01$ after putting $b = 1 \, \mbox{nm}$.

\subsubsection{Monomer and charge distribution}
\begin{figure}[h]
 \begin{center}
     \vspace*{1.0cm}
      \begin{minipage}[c]{15cm}
     \includegraphics[width=15cm]{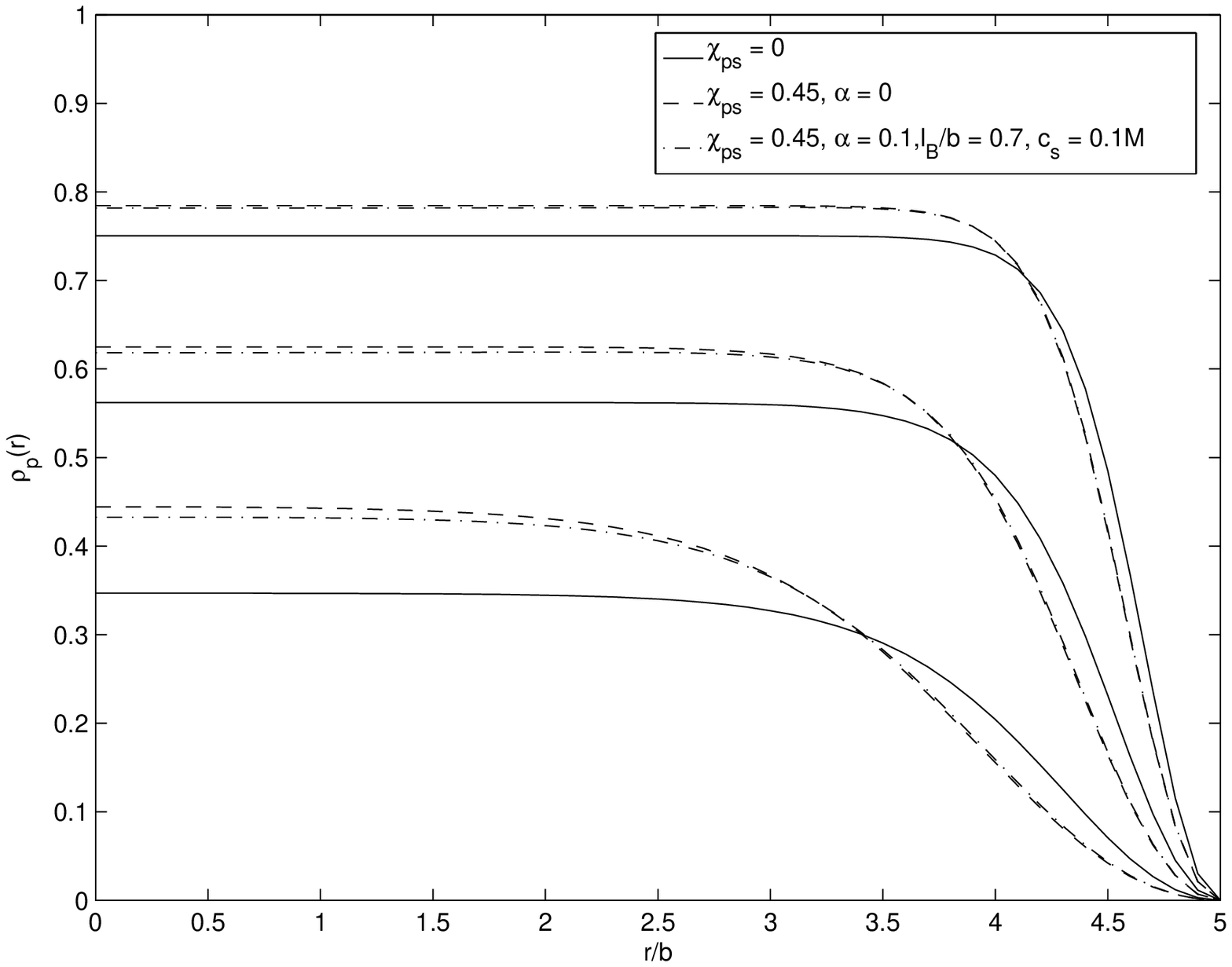}
    \end{minipage}
    \caption{}\label{neffect_expli}
\end{center}
\end{figure}

To study the effect of confinement, we have varied degree of polymerization ($N$) keeping all other
parameters fixed. Increase in $N$ with fixed $R$ leads to a more confined environment with the increase in
monomer density everywhere inside the cavity. This can be seen in Fig. ~\ref{neffect_expli}, where we have plotted monomer densities for a single polyelectrolyte
chain for different values of $N$ after choosing a particular set of parameters ($\chi_{ps} = 0.45,
\alpha = 0.1, l_{B}/b = 0.7, R/b = 5, c_{s} = 0.1 M$). These values for the parameters were chosen to mimic a
salty flexible polyelectrolyte chain under spherical confinement in the presence of water as a solvent.
Also, for comparison purposes,
we have plotted monomer densities for the corresponding neutral chains in
an athermal solvent ($\chi_{ps} = 0$) and in an equivalent
solvent condition ($\chi_{ps} = 0.45$).
Comparing the monomer density for the neutral chain in athermal solvent with
a poorer solvent, it is clear that the exclusion of solvent molecules from
the core of the coil is
stronger as the solvent quality is decreased, as expected.
Now, when the polymer in the less good solvent is charged, the solvent exclusion effect
is slightly weaker. This difference becomes negligible as the monomer volume fraction increases.
Moreover, monomer density profiles for
neutral and the polyelectrolyte chain in the same solvent are almost identical,
which corroborates the coil conformation of the polyelectrolyte.

\begin{figure}[h]
\begin{center}
     \vspace*{1.0cm}
      \begin{minipage}[c]{15cm}
     \includegraphics[width=15cm]{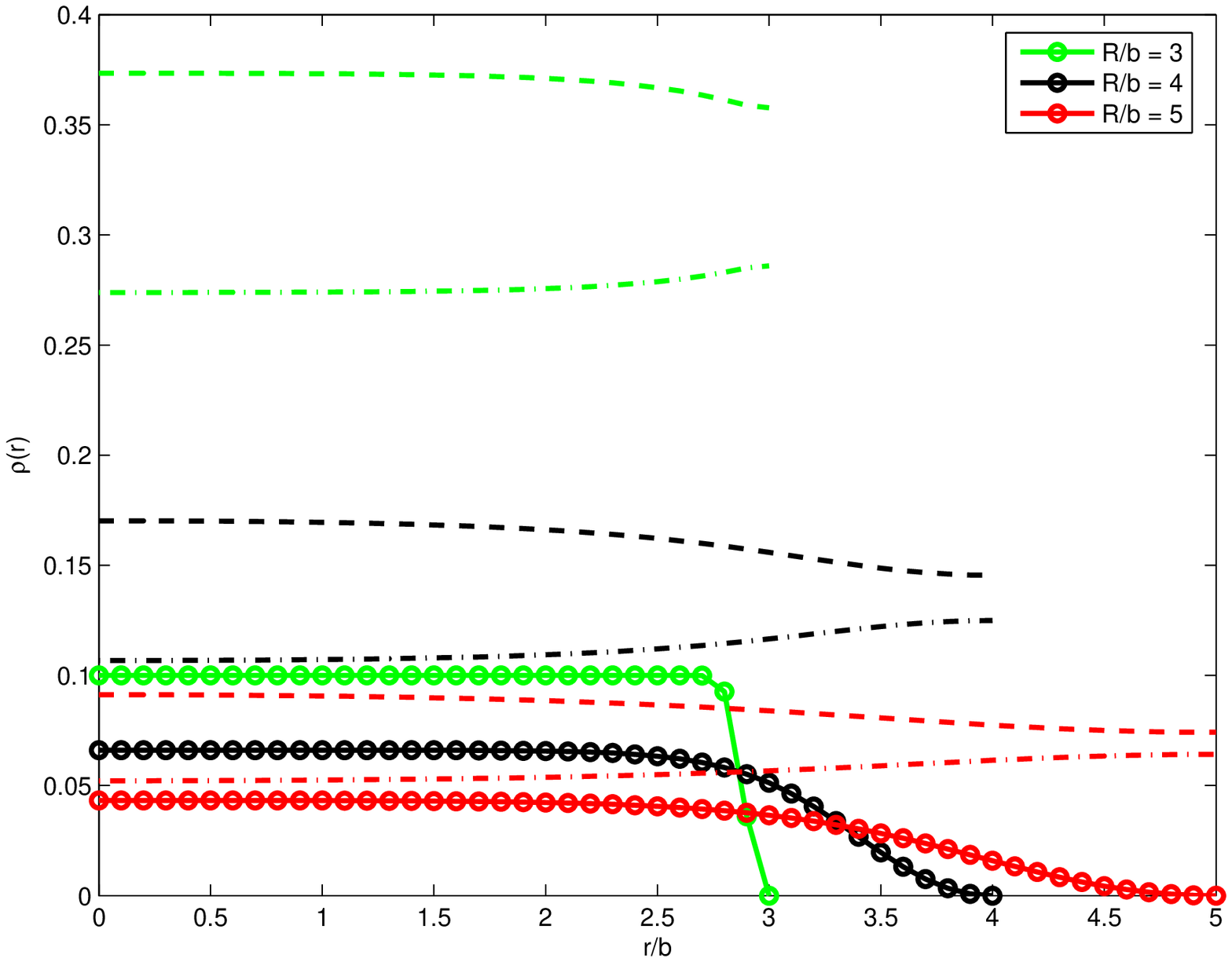}  
    \end{minipage}
\caption{}\label{rcseffect_ion_expli}
\end{center}
\end{figure}

\begin{figure}[h]
\begin{center}
     \vspace*{1.0cm}
      \begin{minipage}[c]{15cm}
     \includegraphics[width=15cm]{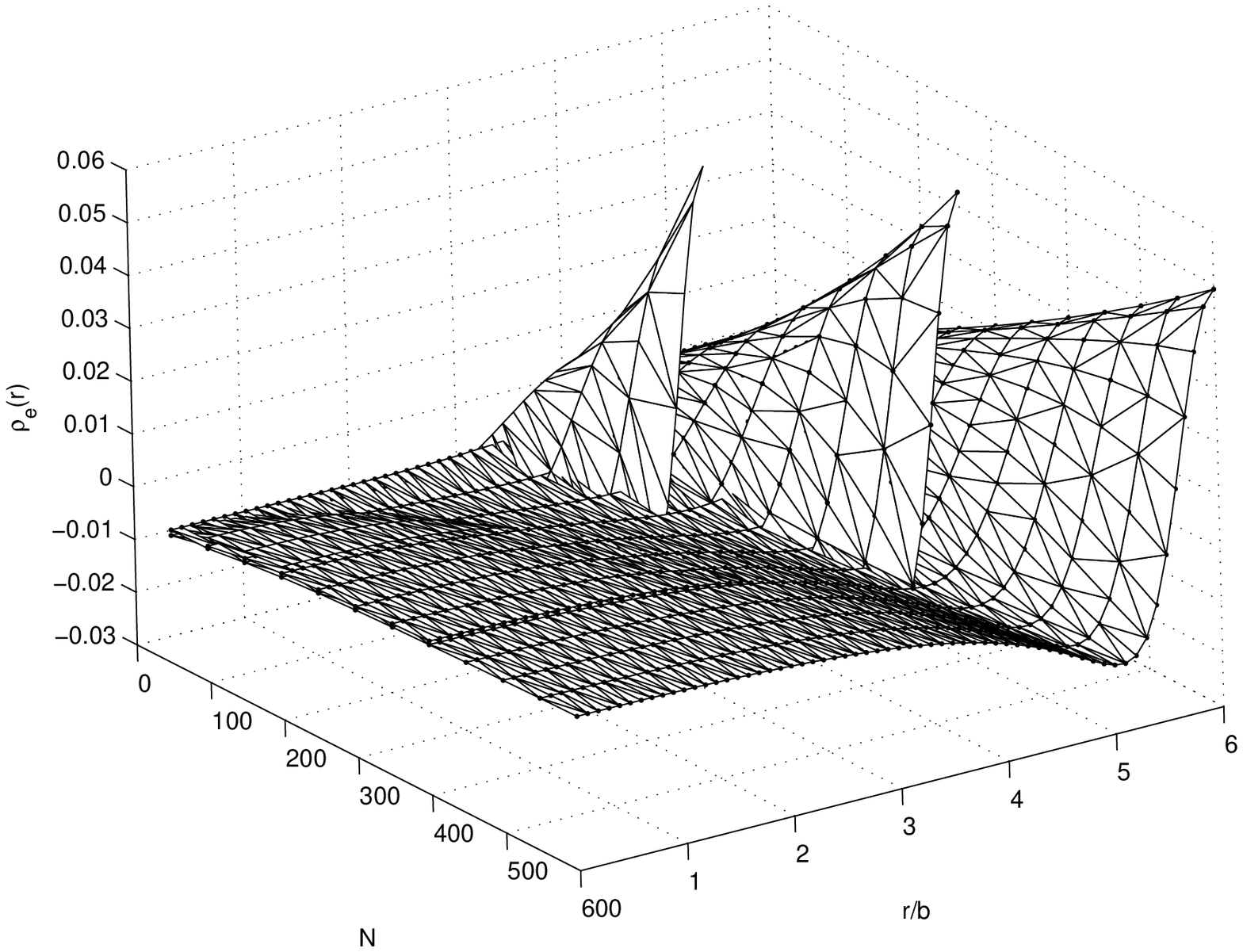}  
    \end{minipage}
\caption{}\label{reffect_charge}
\end{center}
\end{figure}

We have also varied $R$ by keeping the number of salt ions and $\alpha$ fixed,
and the results obtained for radial densities are
shown in Fig. ~\ref{rcseffect_ion_expli} for a particular value of $N$
(number of salt-ions corresponds to $c_{s} = 0.1M$ for $R/b = 5$).
The observed difference in small ion density
profiles for different $R$'s is attributed to
the fact that total salt concentration ($c_{s}$) is changing when $R$ is being changed
with the number of small ions kept fixed during this variation. In this figure,
we have chosen $N=100$ and
the rest of the parameters are the same as in Fig. ~\ref{neffect_expli} for the polyelectrolyte chain.
From Fig. ~\ref{rcseffect_ion_expli}, it is evident that due to
exclusion of the monomers from confining
sphere surface, decrease in $R$ leads to an
increase in concentration of monomers in the interior.
Had it been a bulk situation ($R\rightarrow \infty$), small ion densities would have
reached their bulk value ($c_{s}$).  Looking at small ions' densities
in Fig. ~\ref{rcseffect_ion_expli}, it is clear that for the confined
spherical system, this common presumed result is no more
valid and small ion density profiles haven't reached their bulk value.
In other words, net radial charge density is no more zero near the confining boundary (as shown in Fig. ~\ref{reffect_charge}) and a double layer system\cite{verweybook} is set up as a result of depletion of the chain from the surface.

In this study, small ions are treated as point charges and hence, are not excluded
from the surface of the spherical cavity in contrast to the monomer.
As a result, when $N$ is increased keeping $\alpha$ fixed
at stronger confinements (in terms of higher monomer densities), there are more number
of counterions (positively charged) generated by the chain and
hence, charge density near the surface of the cavity increases (Fig. ~\ref{reffect_charge}).
On the other hand, near the center of the cavity, the local charge density
curves move toward zero as $N$ is increased (i.e. interior of the cavity
is becoming locally electroneutral). For lower degrees of confinements (i.e. low
$\bar{\rho}_{p}$), increase in $N$ leads to increase in charge density everywhere
due to more number of ions in the system. However, this behavior is seen over a very small
density regime.

We have also varied other parameters ($c_{s},\alpha$ and $l_{B}$). Effects of
all these parameters on monomer densities are consistent with
the observation that a polyelectrolyte\cite{muthu96} chain at high salt concentration
is equivalent to a neutral chain with an effective excluded volume parameter given by

\begin{eqnarray}
w_{eff} &=& \frac{1}{1-\bar{\rho}_{p}}- 2\chi_{ps} + \frac{4\pi l_{B}\alpha^{2}Z_{p}^{2}}{\kappa^{2}}. \label{eq:weff}
\end{eqnarray}
So, the polyelectrolyte chain
shows higher expansion\cite{muthubeer97}
as compared to its neutral analog\cite{muthunick87}.

On the other hand, at low salt concentrations,
the chain can tend to attain a rod-like conformation due to dominance of electrostatic repulsions along the
backbone and spherical symmetry is broken. In our mean field study, we use spherical symmetry and
Gaussian model for the polyelectrolyte chain.
Due to these limitations, spherical symmetry breaks
down in the extremely low salt regime and we are not allowed to explore the low
salt regime using the current model.

These results show that confinement of the chain leads to
the development of a charge density wave inside the cavity and
an outcome of this charge density wave is that there is a
potential difference across the center of the cavity and the surface.
It is to be noted that for a confined system with only small ions inside,
potential is constant everywhere and local electroneutrality is the
equilibrium state (trivial solution of SCF equations).
However, it is the depletion of the chain from the cavity surface,
which leads to the accumulation of charges near the surface and as
a result inhomogeneous charge distribution is attained. To study the
role of polyelectrolyte in the free energy, we compare the
free energies of the inhomogeneous phase with the cavity containing pure solvent and salt-ions
in the next section.

\subsubsection{Free energy within saddle point approximation}

\begin{figure}[h]
\begin{center}
     \vspace*{1.0cm}
      \begin{minipage}[c]{15cm}
     \includegraphics[width=15cm]{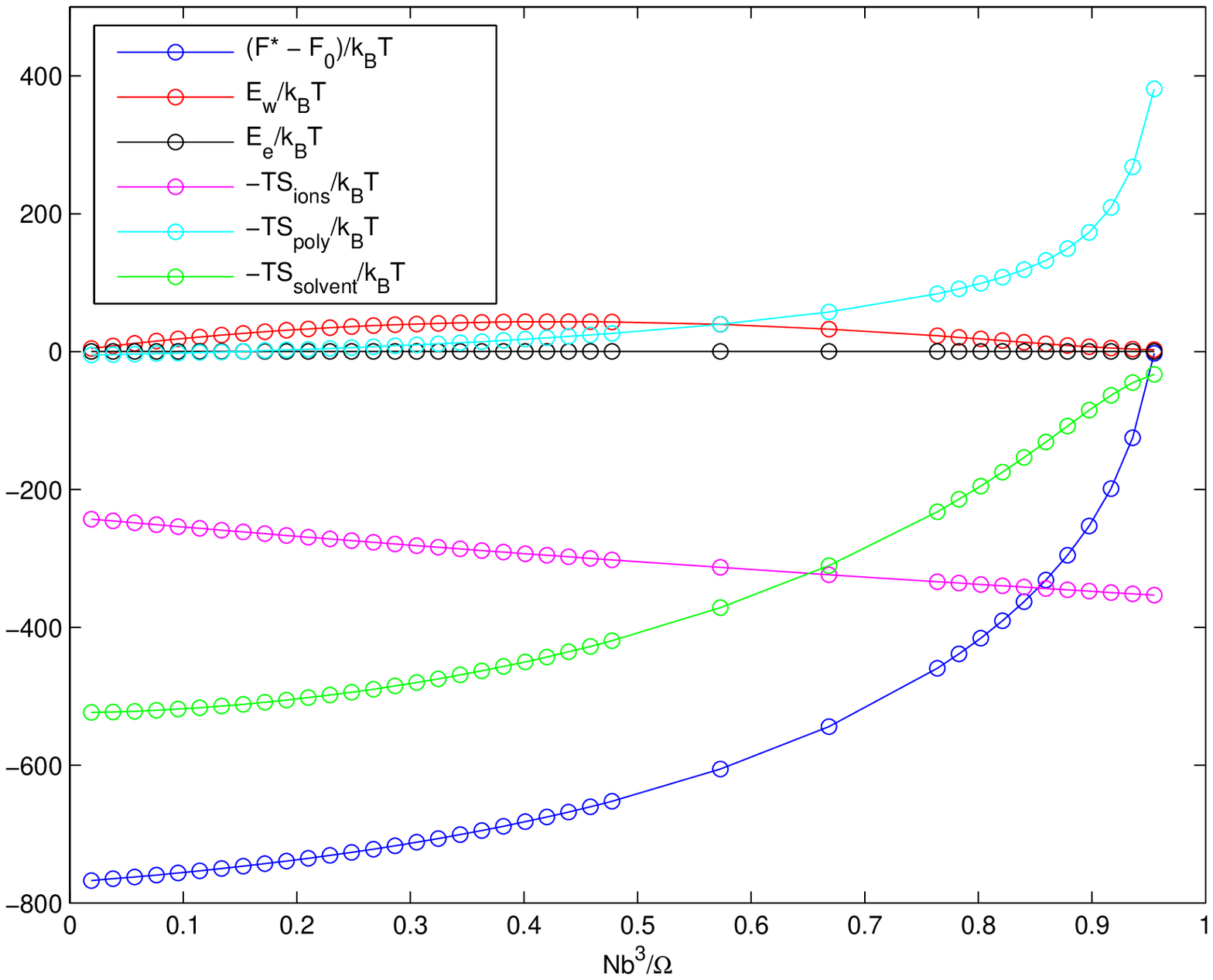}  
    \end{minipage}
\caption{}\label{free_contri_expli}

\end{center}
\end{figure}

In Fig. ~\ref{free_contri_expli}, we have plotted different contributions to
free energy for a salty system using Eqs. (~\ref{eq:free_thermo_exp} - ~\ref{eq:solvent_ent}).
Analyzing the contributions to free energy, it is clear that
the free energy has four major contributions. At lower polymer volume fractions,
free energy is dominated
by entropy of small ions ($-TS_{ions}/k_{B}T$) and solvent ($-TS_{solvent}/k_{B}T$).
As volume fraction is increased, chain conformational entropy
and polymer-solvent interaction energy also become important. Also, electrostatic energy
part in free energy
is small compared to other terms,
stressing a minor role played by electrostatic energy in the crowded environment
under investigation here. Share of each contribution to
the free energy depends on the degree of confinement. For instance,
when degree of confinement is extremely
high ($Nb^{3}/\Omega\rightarrow 1$), solvent
entropy and polymer-solvent interaction energy terms are
negligible and free energy has two
major contributions - small ion entropy and chain conformational entropy.
Strictly speaking, our theoretical model breaks down as soon as monomer density inside the cavity
is close to unity because the finite size of small ions and nature of interactions between the
various species become important. All these effects can not be captured with our theory,
 which involves only two body interaction
potential.

A few comments about the shape of the plots in
Fig. ~\ref{free_contri_expli} are due here. Shape of small ions entropy term is
governed by number of small ions term ($\sum_{j}\int d\mathbf{r}\rho_{j}(\mathbf{r})$  in
Eq. (~\ref{eq:tsmall})) and when $N$ is increased keeping degree of
ionization ($\alpha$) and $R$ fixed, number of
counterions increases and hence, $-TS_{ions}/k_{B}T$ decreases with increase in $N$
 (almost linearly).
Similarly, when $N$ is increased while keeping $R$ fixed, total number
of solvent molecules in the cavity decreases (due to incompressibility) and hence, $-TS_{solvent}/k_{B}T$
increases.
Increase in $N$ for a fixed $R/b$ leads to lower number of
conformations available to the chain and hence,
entropy of the chain decreases or $-TS_{poly}/k_{B}T$ increases.
Shape of excluded volume interaction energy and electrostatic
interaction energy
can be understood by the fact that in the asymptotic
limit with respect to degree of confinement,
the product of densities (monomer and solvent) as well as the electric potential are
small.


In implicit solvent computations\cite{muthujoey04}, free energy goes linearly with $N^{2}/\Omega$.
In order to see whether the same linear law is followed by free energy in
explicit solvent model, we have plotted free energy as a function of
$N^{2}/\Omega$ for different values of $R/b$ in Fig. ~\ref{free_vs_n2r3}.
It is found that free energy follows the linear law only for
lower values of $\bar{\rho}_{p}$. For higher $\bar{\rho}_{p}$, there are
deviations from
this linear law due to the conformational
entropy of the chain. Overall, the shape of the free energy curve is
of the form $\sum_{j=c,+,-,s}n_{j}\left[\ln (n_{j}/\Omega) -1\right]$
for lower $\bar{\rho}_{p}$ and systematic deviations are seen for higher
$\bar{\rho}_{p}$.

Also, to highlight the role of the polyelectrolyte in the free energy of
confinement, we have plotted the \textit{difference} between the free energy of the spherical
cavity with and without chain (i.e. $\Delta F^{\star} = F^{\star} - F\{\bar{\rho}_{p} = 0\}$)
for different radii of the confining cavity and monomer densities in Fig. ~\ref{free_delta_salt_expli}. In these computations, it has been assumed that $w_{pp} = w_{ss}$ to get rid of an uninteresting constant. The results reveal that the confinement of the chain is a thermodynamically unfavorable process and for the same density, larger value
 of $\Delta F^{\star}$ for larger spherical cavity can be attributed to the difference in conformational entropy and polymer-solvent interaction energy for chains of different lengths in spheres of different radii. 

 In scaling theories\cite{scalingbook}, it is commonly asserted that the confinement free energy is extensive in $N$. In order to confirm this assertion, we have plotted $(F^{\star} - F_{0})/Nk_{B}T$ for various values of monomer densities at different values of $R$ (see Fig. ~\ref{freebyn_vs_den}). Our results clearly show that the free energy under spherical confinement is not extensive in $N$, in contrast to the rectangular and cylindrical confinements\cite{scalingbook}.

\subsubsection{Osmotic pressure and mean activity coefficient}
 In Fig. ~\ref{osmotic}, we have plotted the osmotic pressure obtained from Eq. (~\ref{eq:osmotic}) as a function of
 monomer density. For comparison purposes, we have also plotted the osmotic pressure of the homogeneous phase (Eq. (~\ref{eq:osmotic_homo})). It is found that the osmotic pressure is the same for the homogeneous and
 the inhomogeneous phase at lower
 monomer densities and ideal gas law for osmotic pressure is obtained in this regime.
 However, at higher densities, deviations from ideal gas law are seen and these deviations are larger for
 the inhomogeneous phase in comparison with the homogeneous phase. Comparing each term in Eqs. (~\ref{eq:osmotic})
 and (~\ref{eq:osmotic_homo}), it is found that the discrepancy between the inhomogeneous and homogeneous phase
 arises as a result of the depletion of the chain from the spherical surface. Due to the depletion, the monomer
 density at one Kuhn step away from the surface is higher for the inhomogeneous system in comparison with the homogenous system. Moreover, the log term involving solvent density and quadratic excluded volume interaction term involving the chi parameter add to the discrepancy. Other than these terms, electrostatic and small ions terms do not change much. 
 Effect of the cavity radius on the osmotic pressure profiles can be easily explained using Eq. (~\ref{eq:osmotic_homo}) and the fact that number densities of salt ions is higher for smaller cavity, when number of ions is kept fixed during the computation.

 In the absence of the polyelectrolyte, the local electroneutrality is the equilibrium state for the small ions. That means the mean activity coefficient for the salt is unity. However, due to the depletion of the chain from the surface, local electroneutrality gets broken and that leads to a deviation from unity. In Fig. ~\ref{activity},
 we have plotted the mean activity coefficient for the monovalent salt and it is clear that the presence of the polyelectrolyte chain leads to deviation from ideal behavior (local electroneutrality). The shape of the activity coefficient curves can be understood by the fact that local electroneutrality is attained for very large densities also, where number of small ions is large and almost uniformly distributed. Effect of $R$ on these plots
 can be explained by the fact that the extent of inhomogeneity in electrostatic potential increases with an increase in $R$.

\subsubsection{One-loop fluctuation corrections : narrow depletion zone approximation}
The saddle point approximation used in computing free energy (Eq. (~\ref{eq:free_energy})) is valid when the number
densities of small molecules and monomers is high (i.e. concentrated regime). In order to capture the role of fluctuations in the concentrated regime, we have expanded
the integrand of functional integrals over fields around the saddle point solution up to quadratic terms in fields (Appendix-B) so that the functional
integrals to be carried out are Gaussian (one-loop calculations). For low concentrations (dilute and semi-dilute regimes), this treatment
breaks down and other techniques have to be employed.

Even at the one-loop level, it is very difficult to sum the infinite series,
which emerge as a result of Gaussian integrals . Moreover, these sums are plagued
with ultraviolet divergences, which have to be regularized. However, we have been
able to sum these series in the long chain limit ($N\rightarrow \infty$)
when $\kappa R \rightarrow \infty$ and $\xi^{-1} R \rightarrow \infty$, where $\kappa^{-1}$ and $\xi$ are Debye and Edwards' screening lengths, respectively.  In the concentrated regime for a very long chain, the width of the depletion zone near the surface of the sphere is very small as compared to the radius of the sphere and we can suppress the radial dependence of the densities (cf. Fig. ~\ref{neffect_expli} and ~\ref{rcseffect_ion_expli}). Taking this approximation and ignoring the correlation energy of the charges along the backbone of the chain (which is very small for a weakly charged polyelectrolyte), the
free energy at one-loop level can be written as (Appendix-B)

\begin{eqnarray}
\frac{F}{k_{B}T}&=& \frac{F^{\star}}{k_{B}T} +  \frac{1}{2}\sum_{k=1}^{\infty}\sum_{l=0}^{\infty}(2l + 1)\left[\ln\left(1 + \frac{R^{2}/\xi^{2}}{\nu_{kl}^{2}}\right)+ \ln\left(1 + \frac{\kappa^{2}R^{2}}{\nu_{kl}^{2}}\right)\right ], \label{eq:free_approx_sum}
\end{eqnarray}
where $\kappa^{2} = 4\pi l_{B}\sum_{c,+,-}Z_{j}^{2}n_{j}/\Omega , \xi^{-2} = 12 (\frac{1}{1-\bar{\rho}_{p}} - 2\chi_{ps}) N b/\Omega$ and $\nu_{kl}$ is $k^{th}$ zero of the spherical Bessel function of order $l$ (i.e. $j_{l}(\nu_{kl}) = 0$). The infinite sum over $k$ can be computed exactly. However, the sum over $l$ diverges. The divergence can be regularized by introducing
an upper cutoff $M$ on $l$ and identifying cut-off independent part. The \textit{finite} sum over $l$ has been computed in Ref. \cite{finite_tellez}. The cut-off independent part gives

\begin{eqnarray}
\frac{F}{k_{B}T}&=& \frac{F^{\star}}{k_{B}T} -  \frac{(\kappa^{3} + \xi^{-3})\Omega}{12\pi} + \frac{\kappa^{2}S}{32\pi}\left( 1 +  2 \ln \kappa R\right ) + \frac{\xi^{-2}S}{32\pi}\left( 1 +  2 \ln \xi^{-1} R\right ) + \frac{1}{3}(\kappa + \xi^{-1})R, \nonumber  \\
&& \label{eq:free_fluctuations}
\end{eqnarray}
where $S$ is the surface area of the sphere ($= 4\pi R^{2}$). In the above expression, cubic terms in $\kappa$ and $\xi^{-1}$
 represent the bulk contribution of fluctuations to the free energy and other terms correspond to the finite size
 corrections, which arise as a result of confinement. Note that for $R\rightarrow \infty$, these finite size contributions vanish and the well-known screening result is obtained. Also, the finite size contributions are smaller than the
 bulk contributions in the limit discussed here and hence, overall, fluctuations lower the total free energy.

Fluctuation corrections to the osmotic pressure and the mean activity coefficient can be estimated in a straightforward way by using Eq. (~\ref{eq:free_fluctuations}). It must be kept in mind that Eq. (~\ref{eq:free_fluctuations})
is strictly valid in the strong screening regime for an infinitely long chain. In order for these
conditions to be realized, spherical cavity has to be very large to accommodate the long chain (because of incompressibility condition). On the other hand, numerical solution of SCF equations for large spherical cavities with a very long chain becomes very expensive. So, in this study, we limit ourselves to a qualitative discussion about the fluctuation corrections to osmotic pressure and mean activity coefficient.

Qualitatively, the osmotic pressure is decreased due to fluctuations (because the fluctuations at one loop level
lower the free energy). It can be shown quite easily that the leading corrections to the pressure profiles\cite{edwardsbook,mcquarie} will be of the form $-\kappa^{3}/24\pi$ and $-\xi^{-3}/24\pi$. In the thermodynamic limit of infinite volume for an infinitely long chain, these expressions become exact. As the osmotic pressure must be positive, this analysis sets the range of validity of the one-loop calculations\cite{edwardsbook} to be well above the overlap concentration (in the concentrated regime). In fact, below those concentrations, perturbative treatment
to capture the role of fluctuations fails and non-perturbative methods to capture the role of higher order terms in the expansion of the integrand have to be used (e.g. in dilute and semi-dilute regime)\cite{muthu96,edwardsbook}.

The finite size corrections to these laws (originating from the last three terms on the right hand side in Eq. ~\ref{eq:free_fluctuations}) are small but positive and overall, increase the range of validity of the fluctuation analysis. Similar analysis can be carried out for the mean-activity coefficient of the monovalent salt and the leading corrections coming from fluctuations\cite{mcquarie} are of the form $\mbox{exp}\left(-\kappa l_{B}/2\right)$. Of course, the analysis is valid for the low salt concentrations (Debye-H\"{u}ckel regime) because the size and the nature of the short-range interactions of salt ions become important at higher concentrations.

\section{Conclusions}
\setcounter {equation} {0} \label{sec:conclusions}
We have studied a weakly charged flexible polyelectrolyte chain
under spherical confinement using SCFT.
In Sections ~\ref{sec:theory} and ~\ref{sec:results},
we have demonstrated that SCFT predicts creation of a charge density wave
and a potential difference across the center of the sphere and the boundary.

We have also shown that for a given
charge density along the backbone, free energy of a
 flexible chain ($F^{\star} - F_{0}$) has four major contributions - entropy of small ions,
entropy of solvent , energy due to polymer-solvent interactions
and conformational entropy of the chain. Share of each contribution to the free energy depends on
the monomer density, degree of ionization and salt-concentration inside the sphere.
However, electrostatic interaction energy plays a minor role in free energy for
the weakly charged flexible polyelectrolyte. Our results show that the free energy
is not extensive in the number of monomers.

Osmotic pressure for the polyelectrolyte chain follows ideal gas law in the low monomer density
regime and substantial deviations are seen as the monomer density is increased.
Mean activity coefficient for the monovalent salt show a small, yet systematic deviation
from unity, highlighting the role of the polyelectrolyte in breaking
the local electroneutrality condition seen in the absence of the polyelectrolyte.

One-loop fluctuation analysis (within the approximation of narrow depletion zone)
reveals that fluctuations lower the free energy and the
free energy has additional contributions coming from
screening effects due to monomers as well as small-ions. Without any confinement, the fluctuations
corrections at one-loop level\cite{edwardsbook} have been shown to lower the free energy
and the range of validity of the fluctuation analysis
is set by the condition that osmotic pressure must be positive.
The concentrations over which the condition is satisfied, come out to be well-above the
overlap concentration in polymer solution theories. As the finite size corrections
to the fluctuation contributions in the free energy are positive, so the range of
validity of the fluctuation analysis gets widened due to the confinement.

Finally, we comment on the assumptions used in arriving at the above conclusions:\\
(1) We have used the spherical symmetry of
the system which is valid as long as coil conformation of the chain is
retained and a rod like
conformation is avoided. Also, we have used SCFT for a flexible polyelectrolyte chain,
which means
the chain must not be stiffened due to charges on the backbone and large torsional barriers.
These criteria are realized in the presence of high salt,
where electrostatic interactions get screened and become short ranged. \\

(2) Saddle-point approximation breaks down in the extremely dilute limit for monomers,
solvent and small-ions. The approximation is strictly valid in the concentrated regime
in the presence of enough salt and solvent. The necessity to have many solvent molecules, sets
upper limit to be away from extremely dense regime, where $Nb^{3}/\Omega \rightarrow 1$.

In this study, fluctuations in densities and fields about the mean field
solution of SCF equations have been treated perturbatively by expanding free energy functional up to
quadratic order about the saddle point. This fluctuation
analysis breaks down for low density regime, where higher order terms also contribute and need to be taken into
account using an appropriate non-perturbative treatment. \\

(3) Role of solvent is taken into account by taking
the volume of a solvent molecule as the same as that of a monomer. In reality, solvent size may be smaller
than Kuhn step length and incompressibility
constraint is violated near the boundary. For the present study, we have
simply assumed that the effect of the depletion zone for solvent on the system properties
is negligible and can be ignored. All these limitations of the current
model can be removed by taking the solvent size to be different from the monomer\cite{matsen02}
and using $\rho_{p}(r) + \rho_{s}(r) = \rho(r)$, where $\rho(r)$ is a
suitable function, which is $\rho_{0}$ away from the boundary and
falls from $\rho_{0}$ to zero in a smooth fashion within the depletion
zone\cite{matsen02}. However, the choice of $\rho(r)$ and width of depletion zone is
arbitrary and depends on the numerics
of the problem.

Similarly, small ions have been treated as point charges, which might not be a bad approximation knowing the fact that
typical ion radii for monovalent cations and anions lie in the range $\sim 0.06 - 0.22$ nm\cite{intermolecular_book}.\\

(4) In principle, the degree of ionization ($\alpha$) should also be computed by minimization of
free energy\cite{muthu04} with respect to $\alpha$.
In order to simplify the numerical work, we have taken the degree of ionization ($\alpha$)
to be independent of $l_{B}$ and avoided any ion condensation effects.\\

(5) While splitting the free energy into energy and entropy, we have assumed that dielectric
constant ($\epsilon$) of the solvent is insensitive to temperature. However, the temperature
dependence of the dielectric constant\cite{marcus55,overbeek90} can be easily incorporated in our theory. \\

\section*{ACKNOWLEDGEMENT}
\setcounter {equation} {0} \label{acknowledgement}
 We acknowledge
financial support from NIH Grant 5R01HG002776-05, NSF Grant No. DMR 0605833,
and the Material Research Science and Engineering Centre at the University of
Massachusetts, Amherst.

\renewcommand{\theequation}{A-\arabic{equation}}
  \setcounter{equation}{0}  
  \section*{APPENDIX A : SCFT with explicit solvent }
Here, we present a summary of the steps in obtaining the saddle point equations for
a single polyelectrolyte chain in the presence of salt ions and solvent molecules (section (\ref{sec:theory_scf}) ). The procedure is similar to the ones presented in Refs. \cite{shi,wang}. As the first step, we define microscopic densities as
 \begin{eqnarray}
    \hat{\rho}_{p}(\mathbf{r})  &=& \frac{1}{b}\int_{0}^{Nb} ds \, \delta (\mathbf{r}-\mathbf{R}(s)) \\
     \hat{\rho}_{j}(\mathbf{r}) &=&  \sum_{i=1}^{n_{j}} \delta (\mathbf{r}-\mathbf{r}_{i}) \quad \mbox{for} \quad j = s,c,+,-\\
\hat{\rho}_{e}(\mathbf{r}) &=& e \left [ \alpha Z_{p} \hat{\rho}_{p}(\mathbf{r}) + \sum_{j = c,+,-}Z_{j}\hat{\rho}_{j}(\mathbf{r}) \right ],
\end{eqnarray}
where $\hat{\rho}_{p}(\mathbf{r}), \hat{\rho}_{j}(\mathbf{r})$ and $\hat{\rho}_{e}(\mathbf{r})$
stand for monomer, small molecules (ions and solvent molecules) and local charge density, respectively.

As the second step, we use a functional integral representation for the incompressibility constraint
\begin{eqnarray}
\prod_{r} \delta\left(\hat{\rho}_{p}(\mathbf{r}) + \hat{\rho}_{s}(\mathbf{r}) - \rho_{0}  \right)
 &=& \int D[w_{+}(\mathbf{r})]e^{-i\int d\mathbf{r} w_{+}(\mathbf{r})\left(\hat{\rho}_{p}(\mathbf{r}) +
\hat{\rho}_{s}(\mathbf{r}) - \rho_{0}        \right)},
 \end{eqnarray}
where $w_{+}(\mathbf{r})$ is the well-known pressure field which enforces the incompressibility
constraint at all points in the system and $i = \sqrt{-1}$.

As the third step, dimensionless Flory's chi parameter is introduced to
club together the three excluded volume parameters so that
\begin{eqnarray}
       \chi_{ps}b^{3} & = & w_{ps} - \frac{w_{pp} + w_{ss}}{2}.
\end{eqnarray}
Note that the clubbing of excluded volume parameters has been possible because of
the incompressibility constraint so that only one independent parameter appears in the
theory. Otherwise, there would have been three independent
parameters.

Following these three steps, Eq. (~\ref{eq:parti_sing}) becomes
\begin{eqnarray}
       \mbox{exp}\left(-\frac{F} {k_{B}T}\right )& = & \frac {1}{\prod_{j}n_{j}!}\int D[\mathbf{R}] \int \prod_{j} \prod_{m=1}^{n_{j}} d\mathbf{r}_{m} \int D[w_{+}(\mathbf{r})]\quad \mbox{exp} \left \{-\frac {3}{2 b}\int_{0}^{Nb}dt\left(\frac{\partial \mathbf{R}(t)}{\partial t} \right )^{2}
\right . \nonumber \\
&& - i \int d\mathbf{r} w_{+}(\mathbf{r})\left(\hat{\rho}_{p}(\mathbf{r}) +
\hat{\rho}_{s}(\mathbf{r}) - \rho_{0} \right) - \chi_{ps}b^{3}\int d\mathbf{r} \hat{\rho}_{p}(\mathbf{r})\hat{\rho}_{s}(\mathbf{r}) \nonumber \\
&& \left .
- \frac{1}{2}\int d\mathbf{r}\int d\mathbf{r}\,'\frac{\hat{\rho}_{e}(\mathbf{r})\hat{\rho}_{e}(\mathbf{r}\,')}{\epsilon k_{B}T |\mathbf{r}-\mathbf{r}\,'|}  \right \} \quad \mbox{exp} \left \{ - \frac{\rho_{0}}{2} \left( Nw_{pp} + n_{s}w_{ss}\right )\right \}. \label{eq:parti_den}
\end{eqnarray}

So far, we have written the partition function in terms of the local densities. Now, we want to write this partition function in terms of the order parameter of the system. There are many different choices we can make for the order parameter\cite{ohta86,fredbook}.
However, in this study, we follow the method as
described in ref. \cite{fredbook} to introduce the order parameter. As the fourth step, we define
\begin{eqnarray}
  \hat{\rho}_{+}(\mathbf{r}) &=& \hat{\rho}_{p}(\mathbf{r}) + \hat{\rho}_{s}(\mathbf{r}) = \rho_{0}  \\
 \hat{\rho}_{-}(\mathbf{r}) &=& \hat{\rho}_{p}(\mathbf{r}) - \hat{\rho}_{s}(\mathbf{r}),
\end{eqnarray}
where $\hat{\rho}_{-}(\mathbf{r})$ is the order parameter for the inhomogeneous system.

As the fifth step, to go from densities to fields, we use the Hubbard-Stratonovich
transformation for short range excluded volume interactions as well as long
range electrostatic interactions, so that
\begin{eqnarray}
\mbox{exp}\left(\frac{\chi_{ps}b^{3}}{4}\int d\mathbf{r} \hat{\rho}_{-}^{2}(\mathbf{r})\right  ) &=& \frac{1}{\mu_{-}}\int
D[w_{-}(\mathbf{r})]\mbox{exp}\left [  \int d\mathbf{r} \left \{ w_{-}(\mathbf{r})\hat{\rho}_{-}(\mathbf{r})
- \frac{1}{\chi_{ps}b^{3}}w_{-}^{2}(\mathbf{r})\right \}
\right ], \nonumber \\
&&
\end{eqnarray}

\begin{eqnarray}
\mbox{exp}\left(-\frac{1}{2}\int d\mathbf{r} \int d\mathbf{r}\,'\frac{\hat{\rho}_{e}(\mathbf{r})\hat{\rho}_{e}(\mathbf{r}\,')}{\epsilon k_{B}T |\mathbf{r}-\mathbf{r}\,'|}\right  ) &=& \frac{1}{\mu_{\psi}}\int
D[\psi(\mathbf{r})]\mbox{exp}\left [ - \int d\mathbf{r} \left \{ i \psi(\mathbf{r})\frac{\hat{\rho}_{e}(\mathbf{r})}{e} \right . \right. \nonumber \\
&& \left . \left .\qquad \qquad \qquad \qquad
- \frac{\psi(\mathbf{r})}{8 \pi l_{B}}\bigtriangledown_{\mathbf{r}}^{2}\psi(\mathbf{r})\right \}
\right ],
\end{eqnarray}
where
\begin{eqnarray}
\mu_{-} &=& \int D[w_{-}(\mathbf{r})]\mbox{exp}\left [ -\frac{1}{\chi_{ps}b^{3}}\int d\mathbf{r}         w_{-}^{2}(\mathbf{r})\right ]      \\
\mu_{\psi} &=& \int D[\psi(\mathbf{r})]\mbox{exp}\left [ \frac{1}{8 \pi l_{B}}\int d\mathbf{r}
\psi(\mathbf{r})\bigtriangledown_{\mathbf{r}}^{2}\psi(\mathbf{r})                                       \right ].
\end{eqnarray}

Note that, $w_{+}(\mathbf{r}),w_{-}(\mathbf{r})$ and $\psi(\mathbf{r})$ are the \textit{real} fields,
which can be envisioned as chemical potential fields\cite{helfand75}. Now, taking
these two steps, Eq. (~\ref{eq:parti_den}) becomes
\begin{eqnarray}
 \mbox{exp}\left(-\frac{F} {k_{B}T}\right )& = & \frac{1}{\mu_{-}\mu_{\psi}}
\int D[w_{+}(\mathbf{r})] \int D[w_{-}(\mathbf{r})]\int D[\psi(\mathbf{r})]
\quad \mbox{exp}\left [ - f\left\{w_{+},w_{-},\psi\right \} \right ], \nonumber \\
&& \label{eq:functional_tobe}
\end{eqnarray}
where
\begin{eqnarray}
      f\left\{w_{+},w_{-},\psi\right \} & = & \frac{1}{\chi_{ps}b^{3}}\int d\mathbf{r} w_{-}^{2}(\mathbf{r}) - i \rho_{0}\int d\mathbf{r} w_{+}(\mathbf{r}) - \ln Q_{p}
- \sum_{j=s,c,+,-}n_{j}\ln Q_{j}\nonumber \\
&& - \frac{1}{8\pi l_{B}}\int d\mathbf{r}\psi(\mathbf{r})\bigtriangledown_{\mathbf{r}}^{2}\psi(\mathbf{r})
  + \frac{\rho_{0}}{2} \left( Nw_{pp} + n_{s}w_{ss} + \frac{\chi_{ps}b^{3}}{2}\Omega \rho_{0}\right ) \nonumber \\
 && + \sum_{j=s,c,+,-}\ln n_{j}!
\label{eq:parti_functional}
\end{eqnarray}

\begin{eqnarray}
       Q_{p} & = & \int D[\mathbf{R}(t)]\, \mbox{exp}
\left [ -\frac {1}{b}\int_{0}^{Nb}dt\left \{\frac{3}{2}\left (\frac{\partial \mathbf{R}}{\partial t} \right )^{2}
+  i w_{+}\{\mathbf{R}\} - w_{-}\{\mathbf{R}\} + i Z_{p}\alpha \psi\{\mathbf{R}\} \right \} \right ]  \nonumber \\
&&
\end{eqnarray}
\begin{eqnarray}
Q_{s} & = & \int d\mathbf{r}\quad \mbox{exp} \left [ - \{iw_{+}(\mathbf{r}) + w_{-}(\mathbf{r})\}\right ] \\
Q_{j} & = & \int d\mathbf{r}\quad \mbox{exp} \left [ - iZ_{j}\psi(\mathbf{r})\right ] \quad \mbox{for} \quad j = c,+,-.
\end{eqnarray}

The functional integrations over real fields are to be carried out
by contour integration techniques and are almost impossible to compute exactly.
However, if the number of small molecules is large then we can compute the integrals
in the numerator by steepest descent technique, using the knowledge that
the integral along the constant phase (imaginary part of the integrand) contour
is dominated by the local minima of the integrand\cite{fredbook} (note that the functional
integrals in the denominator are divergent and ignored at the level of
saddle point approximation). So, we approximate the functional integrals
in the numerator by the value of the integrand at the local
minima (where the phase comes out to be zero) so that the final approximated integral is real
( $= f\left\{w_{+}^{\star},w_{-}^{\star},\psi^{\star}\right\}$ ).
However, the saddle point values for $w_{+}$ ($=w_{+}^{\star}$) and $\psi$($=\psi^{\star}$)
come out to be
purely imaginary in contrast to $w_{-}$($=w_{-}^{\star}$), which is real.

Also, it should be noted that the densities remain
unchanged with the shift in fields by an arbitrary constant.
So, we write $iw_{+}(\mathbf{r}) - \frac{1}{2}\chi_{ps}b^{3}\rho_{0} = iw_{+}(\mathbf{r}) $
to get rid of the constant in saddle point equations. Now, deriving local minima equations
for the integrand in Eq. (~\ref{eq:parti_functional})
with respect to $w_{+},w_{-}$ and $\psi$, and using notation
$iw_{+}(\mathbf{r}) - w_{-}(\mathbf{r}) \rightarrow \phi_{p}(\mathbf{r})$,
$iw_{+}(\mathbf{r}) + w_{-}(\mathbf{r}) \rightarrow \phi_{s}(\mathbf{r})$,
$i\psi(\mathbf{r})\rightarrow \psi(\mathbf{r})$, $iw_{+}(\mathbf{r}) \rightarrow \eta(\mathbf{r})$,
Eqs. (~\ref{eq:saddle_exp1}- ~\ref{eq:saddle_last}) are obtained. Using these
saddle point equations and employing Stirling's approximation for $\ln n!$, we obtain
Eq. (~\ref{eq:free_energy}). 

\renewcommand{\theequation}{B-\arabic{equation}}
  \setcounter{equation}{0}  
  \section*{APPENDIX B : Fluctuations around the saddle point }

Here, we provide details for one loop treatment of the fluctuations\cite{freedbook}.
We expand the integrand $f$ in Eq. (~\ref{eq:functional_tobe}) up to
second degree terms in $w_{+},w_{-}$ and $\psi$ around their respective
saddle point values. For convenience in writing the expansion,
we introduce a dummy functional variable $\zeta_{p}(\mathbf{r})$, where $p = 1,2,3$
correspond to $w_{+},w_{-}$ and $\psi$, respectively.
In this notation, the expression for free energy becomes (cf. Eq. (~\ref{eq:functional_tobe}))
\begin{eqnarray}
       \mbox{exp}\left(-\frac{F - F^{\star}} {k_{B}T}\right ) & = & \frac{1}{\mu_{-}\mu_{\psi}}
\int \prod_{p=1}^{3}D[\zeta_{p}(\mathbf{r})] \nonumber \\
&&\mbox{exp}\left [ - \frac{1}{2}\int d\mathbf{r} \int d\mathbf{r}\,' \sum_{m=1}^{3}\sum_{p=1}^{3}
K_{mp}(\mathbf{r},\mathbf{r}\,')(\zeta_{m}(\mathbf{r}) - \zeta_{m}^{\star}(\mathbf{r}\,'))(\zeta_{p}(\mathbf{r}) - \zeta_{p}^{\star}(\mathbf{r}\,'))\right ], \nonumber \\
&& \label{eq:flu_first}
\end{eqnarray}
where
\begin{eqnarray}
K_{mp}(\mathbf{r},\mathbf{r}\,') & = & \frac{\delta^{2}f\left\{\zeta_{1},\zeta_{2},\zeta_{3}\right \}}{\delta \zeta_{m}(\mathbf{r})\delta \zeta_{p}(\mathbf{r}\,')}\mid_{\zeta_{1}^{\star},\zeta_{2}^{\star},\zeta_{3}^{\star}}.
\end{eqnarray}
In the expansion, linear terms in fields vanish because of the saddle point condition. $K_{mp}$ can be computed
in a straightforward way and are presented here for completeness
\begin{eqnarray}
K_{11}(\mathbf{r},\mathbf{r}\,') & = & A(\mathbf{r},\mathbf{r}\,') + B_{s}(\mathbf{r},\mathbf{r}\,') \\
K_{22}(\mathbf{r},\mathbf{r}\,') & = & C(\mathbf{r},\mathbf{r}\,') - A(\mathbf{r},\mathbf{r}\,') - B_{s}(\mathbf{r},\mathbf{r}\,') \\
K_{33}(\mathbf{r},\mathbf{r}\,') & = & -\frac{1}{4\pi l_{B}}\bigtriangledown_{\mathbf{r}}^{2}\delta(\mathbf{r} - \mathbf{r}\,') +
Z_{p}^{2}\alpha^{2}A(\mathbf{r},\mathbf{r}\,') + \sum_{j=c,+,-}Z_{j}^{2}B_{j}(\mathbf{r},\mathbf{r}\,')\\
K_{12}(\mathbf{r},\mathbf{r}\,') & = & K_{21}(\mathbf{r},\mathbf{r}\,') = i\left[A(\mathbf{r},\mathbf{r}\,') - B_{s}(\mathbf{r},\mathbf{r}\,')\right] \\
K_{13}(\mathbf{r},\mathbf{r}\,') & = & K_{31}(\mathbf{r},\mathbf{r}\,') = Z_{p}\alpha A(\mathbf{r},\mathbf{r}\,') \\
K_{23}(\mathbf{r},\mathbf{r}\,') & = & K_{32}(\mathbf{r},\mathbf{r}\,') = i Z_{p}\alpha A(\mathbf{r},\mathbf{r}\,')
\end{eqnarray}
where
\begin{eqnarray}
A\left (\mathbf{r},\mathbf{r}\,'\right ) &=& - \rho_{p}(\mathbf{r}) \rho_{p}(\mathbf{r}\,') + g\left (\mathbf{r},\mathbf{r}\,'\right )\\
g\left (\mathbf{r},\mathbf{r}\,'\right ) &=& \left[\int_{0}^{N}ds' \int_{0}^{s'}ds  q(\mathbf{r},s)G(\mathbf{r},\mathbf{r}\,',s,s')q(\mathbf{r}\,',N-s')\right . \nonumber \\
\quad \quad && \left. + \int_{0}^{N}ds \int_{0}^{s}ds' q(\mathbf{r}\,',s')G(\mathbf{r}\,',\mathbf{r},s',s)q(\mathbf{r},N-s)\right ]\frac{1}{\int d\mathbf{r} \, q(\mathbf{r},N)}  \\
B_{j}\left (\mathbf{r},\mathbf{r}\,'\right ) &=& - \frac{1}{n_{j}}\rho_{j}(\mathbf{r}) \rho_{j}(\mathbf{r}\,') + \rho_{j}(\mathbf{r}) \delta \left(\mathbf{r}-\mathbf{r}\,'\right)\quad \mbox{for} \quad j = s,c,+,- \\
C\left (\mathbf{r},\mathbf{r}\,'\right ) &=&  \frac{2}{\chi_{ps} b^{3}}\delta \left(\mathbf{r}-\mathbf{r}\,'\right).
\end{eqnarray}
In the case of confined chain within a spherical cavity, where electrostatic potential and
monomer density are zero at the boundary (Dirichlet boundary conditions), the functional integral
over $\zeta_{p}$ can be carried out in Eq. (~\ref{eq:flu_first}) by expanding all vectorial quantities in
terms of spherical harmonics ($Y_{lm}$) and spherical Bessel functions ($j_{l}$). For an arbitrary
function $h\left (\mathbf{r},\mathbf{r}\,'\right )$ this means
\begin{eqnarray}
h\left (\mathbf{r},\mathbf{r}\,'\right ) &=& \sum_{k=1}^{\infty}\sum_{l=0}^{\infty}\sum_{m=-l}^{l}h_{kl}d_{kl}(r)d_{kl}(r')Y_{lm}(\theta,\phi)Y_{lm}^{\star}(\theta',\phi')\\
d_{kl}(r) &=& \sqrt{\frac{2}{3}}\frac{j_{l}(\nu_{kl}r/R)}{|{j_{l+1}(\nu_{kl})|}},
\end{eqnarray}
where $\nu_{kl}$ is $k^{th}$ zero of the spherical Bessel function of order $l$ (i.e. $j_{l}(\nu_{kl}) = 0$).
Similarly, we expand $\zeta_{p}(\mathbf{r})$ as
\begin{eqnarray}
\zeta_{p}\left (\mathbf{r}\right ) &=& \sum_{k=1}^{\infty}\sum_{l=0}^{\infty}\sum_{m=-l}^{l}\zeta_{pk}d_{kl}(r)Y_{lm}(\theta,\phi).
\end{eqnarray}
 Now, using orthogonal properties of spherical Bessel and spherical harmonics, the functional integrals can be computed. After some lengthy algebra,
\begin{eqnarray}
\frac{F}{k_{B}T}&=& \frac{F^{\star}}{k_{B}T} + \frac{1}{2}\sum_{k=1}^{\infty}\sum_{l=0}^{\infty}(2l + 1)\ln\left(A_{kl} + B_{s,kl} - 4 \sum_{u=1}^{\infty}\sum_{v=1}^{\infty}A_{ku}C^{-1}_{uv}B_{s,vl}\right) \nonumber \\
&& + \frac{1}{2}\sum_{k=1}^{\infty}\sum_{l=0}^{\infty}(2l + 1)\ln\left(1 + \frac{4\pi l_{B}R^{2}}{\nu_{kl}^{2}}\sum_{j=c,+,-}Z_{j}^{2}B_{j,kl}\right)\nonumber \\
&& + \frac{1}{2}\sum_{k=1}^{\infty}\sum_{l=0}^{\infty}(2l + 1)\ln\left(1 + Z_{p}^{2}\alpha^{2} \frac{ A_{kl}-\sum_{u=1}^{\infty}\sum_{v=1}^{\infty} A_{ku}L^{-1}_{uv}A_{vl}}{\frac{\nu_{kl}^{2}}{4\pi l_{B}R^{2}} + \sum_{j=c,+,-}Z_{j}^{2}B_{j,kl}}\right). \label{eq:free_sum}
\end{eqnarray}

In the above equation, the second term on the r.h.s. represents the contribution of correlations arising due to excluded volume interactions between monomers (neutral polymer contribution), the third term arises as a result of
small-ions density fluctuations and the last term correctly represents the correlation energy of charges along the
backbone of the chain.

Also, $C^{-1}_{kl}$ and $L^{-1}_{kl}$ correspond to the coefficients in the spherical harmonics expansion
for the inverse operator of $C\left (\mathbf{r},\mathbf{r}\,'\right )$ and $L\left (\mathbf{r},\mathbf{r}\,'\right )$, respectively. Inverse operator for any arbitrary functional $h\left (\mathbf{r},\mathbf{r}\,'\right )$ is defined by
\begin{eqnarray}
\int d\mathbf{r}\,'h\left (\mathbf{r},\mathbf{r}\,'\right )h^{-1}\left (\mathbf{r}\,',\mathbf{r}\,''\right ) &=& \delta(\mathbf{r}-\mathbf{r}\,''),
\end{eqnarray}
which gives $\sum_{u=1}^{\infty}h_{ku}h^{-1}_{ul} = \delta_{kl}$, where $\delta_{kl}$ is Kronecker delta.
Also, $L\left (\mathbf{r},\mathbf{r}\,'\right )$ is given by
\begin{eqnarray}
L\left (\mathbf{r},\mathbf{r}\,'\right ) &=& A\left (\mathbf{r},\mathbf{r}\,'\right ) + B\left (\mathbf{r},\mathbf{r}\,'\right ) + 4 \int d\mathbf{r}\,''\int d\mathbf{r}\,'''B\left (\mathbf{r},\mathbf{r}\,''\right )(C - 4B)^{-1}\left (\mathbf{r}\,'',\mathbf{r}\,'''\right )B\left (\mathbf{r}\,''',\mathbf{r}\,'\right ). \nonumber \\
&&
\end{eqnarray}

Unfortunately, computation of the coefficients involved in Eq. (~\ref{eq:free_sum}) requires three dimensional calculations for the densities and it is very hard to compute the sums exactly, which diverge in general.
To gain an insight into the problem, we have identified one particular case, where we can evaluate the first two terms
analytically.

If the number of small molecules (solvent and small ions) is large, then operators
$B_{j}\left (\mathbf{r},\mathbf{r}\,'\right )$ become diagonal and also, number densities ($\rho_{j}(\mathbf{r})$)
show a weak dependence on $\mathbf{r}$. Similarly, if degree of polymerization $N$ is large (strictly if $N\rightarrow \infty$), then monomer density
becomes independent of $\mathbf{r}$ except near the surface of the cavity characterized by the width of the
depletion zone, which can be neglected. If both of these conditions are satisfied, then suppressing position
dependence of densities, we get

\begin{eqnarray}
\frac{1}{2}\sum_{k=1}^{\infty}\sum_{l=0}^{\infty}(2l + 1)\ln\left(A_{kl} + B_{s,kl} - 4 \sum_{u=1}^{\infty}\sum_{v=1}^{\infty}A_{ku}C^{-1}_{uv}B_{s,vl}\right) &=&  \frac{1}{2}\sum_{k=1}^{\infty}\sum_{l=0}^{\infty}(2l + 1)\ln\left(1-\bar{\rho}_{p}\right) \nonumber \\
+ \frac{1}{2}\sum_{k=1}^{\infty}\sum_{l=0}^{\infty}(2l + 1)\ln\left(1 + w_{r}b^{3}A_{kl}\right), &&   \label{eq:neutral_sum}
\end{eqnarray}
where $\bar{\rho}_{p} = \frac{Nb^{3}}{\Omega}$ and $ w_{r} = \frac{1}{1-\bar{\rho}_{p}} - 2\chi_{ps}$
is the renormalized excluded volume interaction parameter.
For constant monomer density (while working in the limit $N\rightarrow \infty$), $A\left (\mathbf{r},\mathbf{r}\,'\right )$ satisfies
\begin{eqnarray}
\bigtriangledown_{\mathbf{r}}^{2}A\left (\mathbf{r},\mathbf{r}\,'\right ) &=& - \frac{12 N}{b^{2}\Omega}\delta (\mathbf{r}-\mathbf{r}\,')
\end{eqnarray}
so that $b^{3}A_{kl} = 12 N b R^{2}/(\nu_{kl}^{2}\Omega)$. Neglecting the correlation energy of the charges along the backbone of the chain (the last term on r.h.s. of Eq. (~\ref{eq:free_sum})) and an ultraviolet divergent sum in Eq. (~\ref{eq:neutral_sum}) (first term on r.h.s. of the equation), Eq. (~\ref{eq:free_approx_sum}) is obtained.
\newpage

\section*{REFERENCES}
\setcounter {equation} {0}
\pagestyle{empty} \label{REFERENCES}

\newpage

\section*{FIGURE CAPTION}
\pagestyle{empty}

\begin{description}
\item[Fig. 1.:] Effect of $N$ on monomer densities - comparison with the corresponding neutral chains . In above plots, we have chosen $R/b = 5$ and $N = 100$, $N = 200$ and $N = 300$ from bottom to top, respectively .
\end{description}

\begin{description}
\item[Fig. 2.:] Effect of cavity radius ($R$) on ion densities. In these plots, we have chosen
 $N = 100, \alpha = 0.1, l_{B}/b = 0.7, \chi_{ps} = 0.45 $ and number of salt ions is kept fixed (in all these plots, number of salt ions is equivalent to salt concentration of $0.1M$ for a sphere of
radius $R/b = 5$). Solid, dashed and dash-dotted lines
represent  $\rho(r) = \alpha \rho_{p}(r), \,\rho_{c}(r) + \rho_{+}(r)$ and $\rho_{-}(r)$, respectively.
\end{description}

\begin{description}
\item[Fig. 3.:] Effect of cavity radius ($R$) on net charge density, $\rho_{e}(r) = \sum_{j=c,+,-}Z_{j}\rho_{j}(r) + Z_{p}\alpha \rho_{p}(r)$.
In these plots, we have chosen $\alpha = 0.1, l_{B}/b = 0.7, \chi_{ps} = 0.45 $ and number of salt ions is kept fixed (in all these plots, number of salt ions is equivalent to salt conc of $0.1M$ for sphere of
radius $R/b = 5$). $N$ is increased in steps of $50$ starting from $50$.
\end{description}

\begin{description}
\item[Fig. 4.:] Different contributions to the free energy within the saddle point approximation for salty systems.
Here, we have chosen $N = 100, \alpha = 0.1, l_{B}/b = 0.7, \chi_{ps} = 0.45, c_{s} = 0.1M, R/b = 5$.
\end{description}

\begin{description}
\item[Fig. 5.:] Effect of confinement on the free energy within the saddle point approximation. Here, we have chosen
 $\alpha = 0.1, l_{B}/b = 0.7, \chi_{ps} = 0.45 $ and number of salt ions is kept fixed so that the number of salt ions is equivalent to salt concentration of $0.1M$ for sphere of
radius $R/b = 5$.
\end{description}

\begin{figure}[h]
\begin{center}
     \vspace*{1.0cm}
      \begin{minipage}[c]{15cm}
\includegraphics[width=15cm]{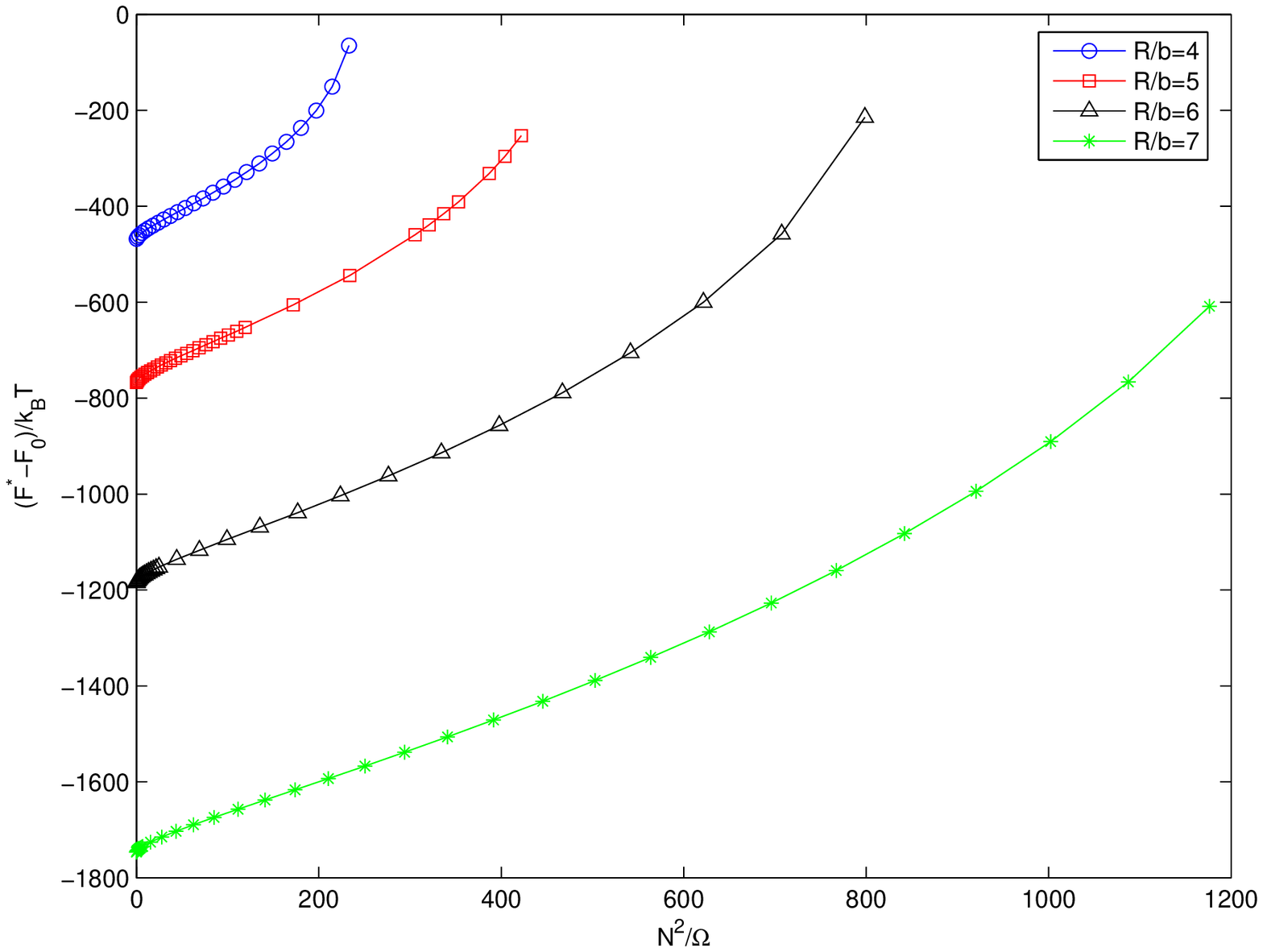}
    \end{minipage}
\caption{} \label{free_vs_n2r3}

\end{center}
\end{figure}

\begin{figure}[h]
\begin{center}
     \vspace*{1.0cm}
      \begin{minipage}[c]{15cm}
     \includegraphics[width=15cm]{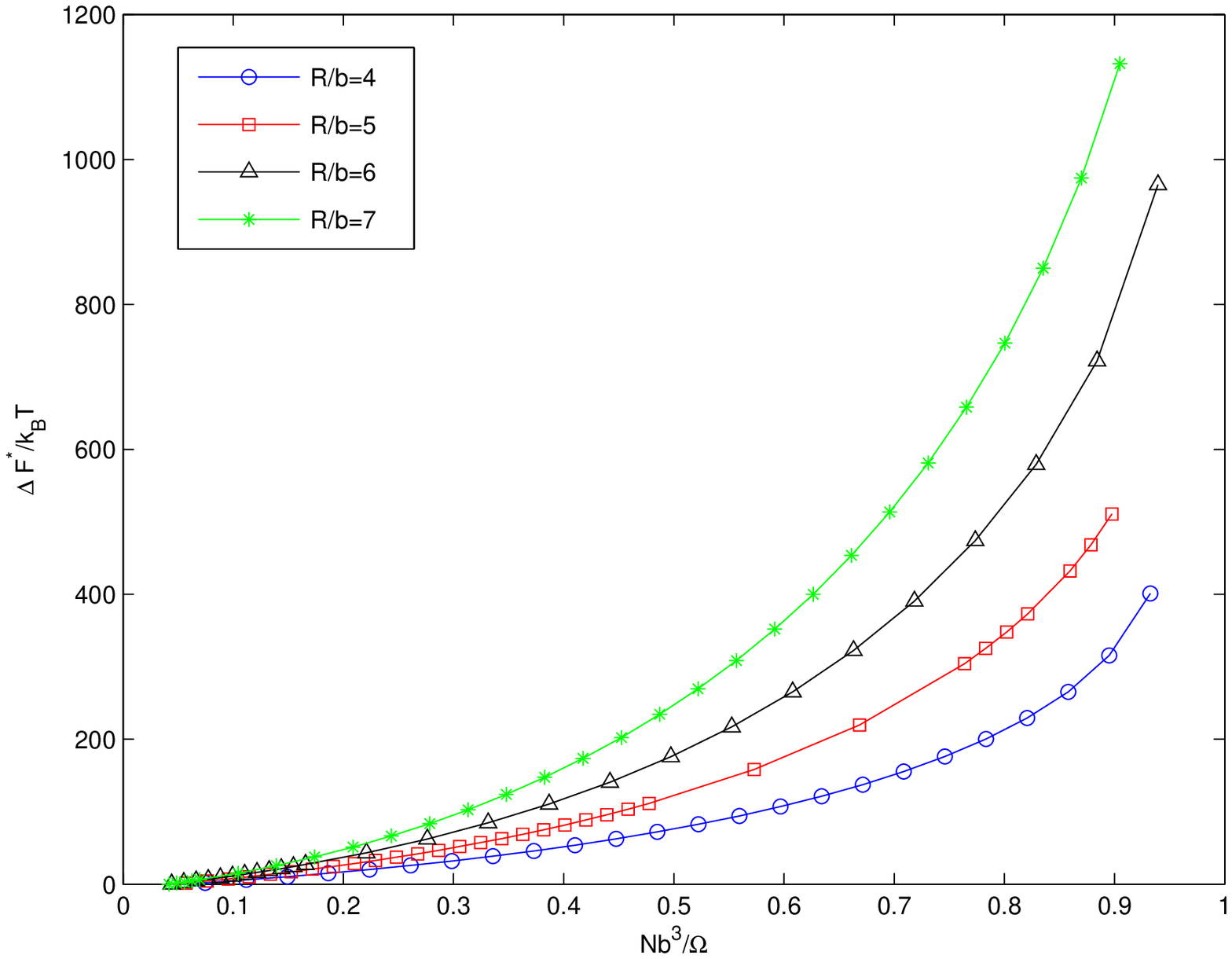}  
    \end{minipage}
\caption{}\label{free_delta_salt_expli}
\end{center}
\end{figure}

\begin{figure}[h]
\begin{center}
     \vspace*{1.0cm}
      \begin{minipage}[c]{15cm}
\includegraphics[width=15cm]{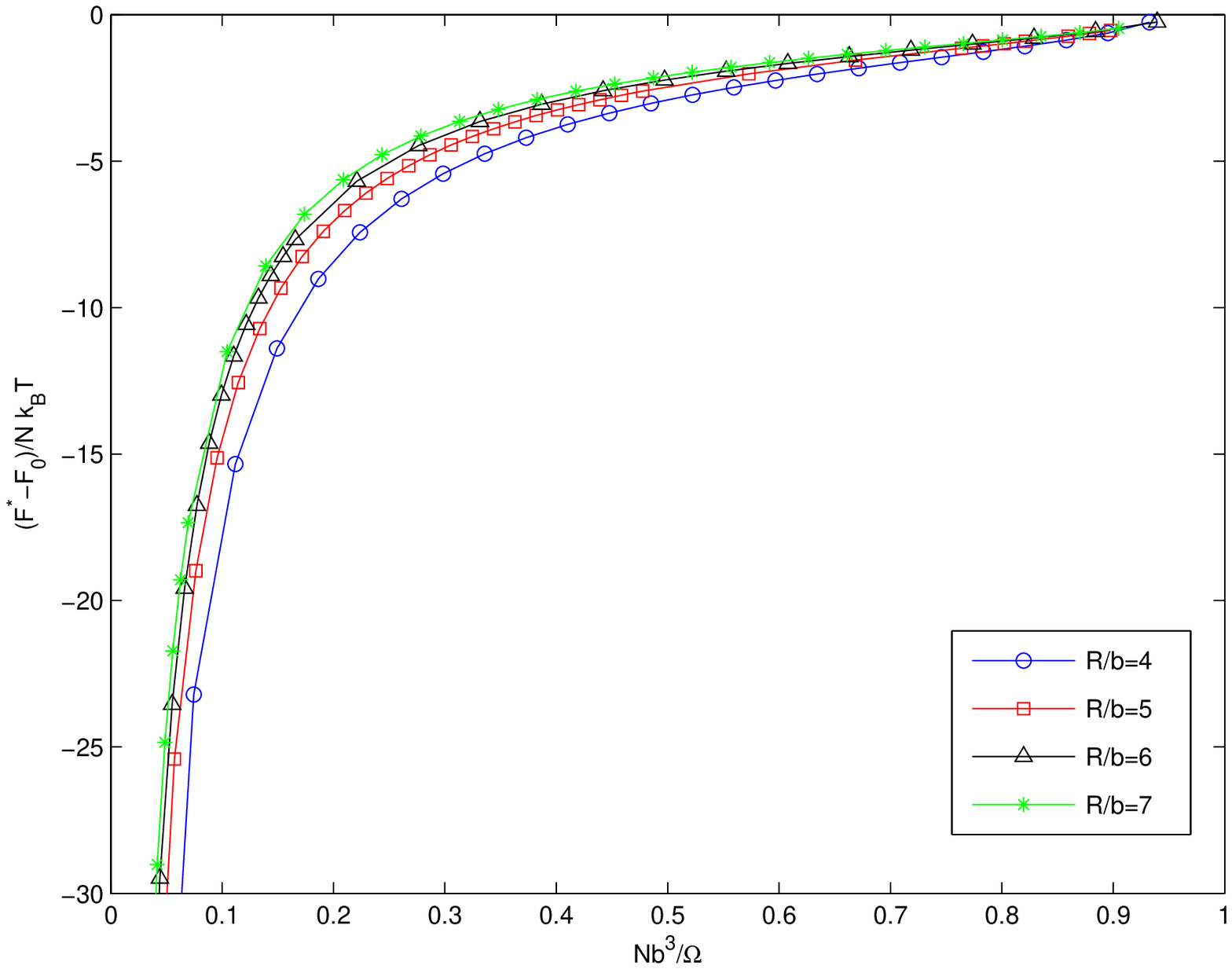}
    \end{minipage}
\caption{} \label{freebyn_vs_den}

\end{center}
\end{figure}

\begin{figure}[h]
\begin{center}
     \vspace*{1.0cm}
      \begin{minipage}[c]{15cm}
     \includegraphics[width=15cm]{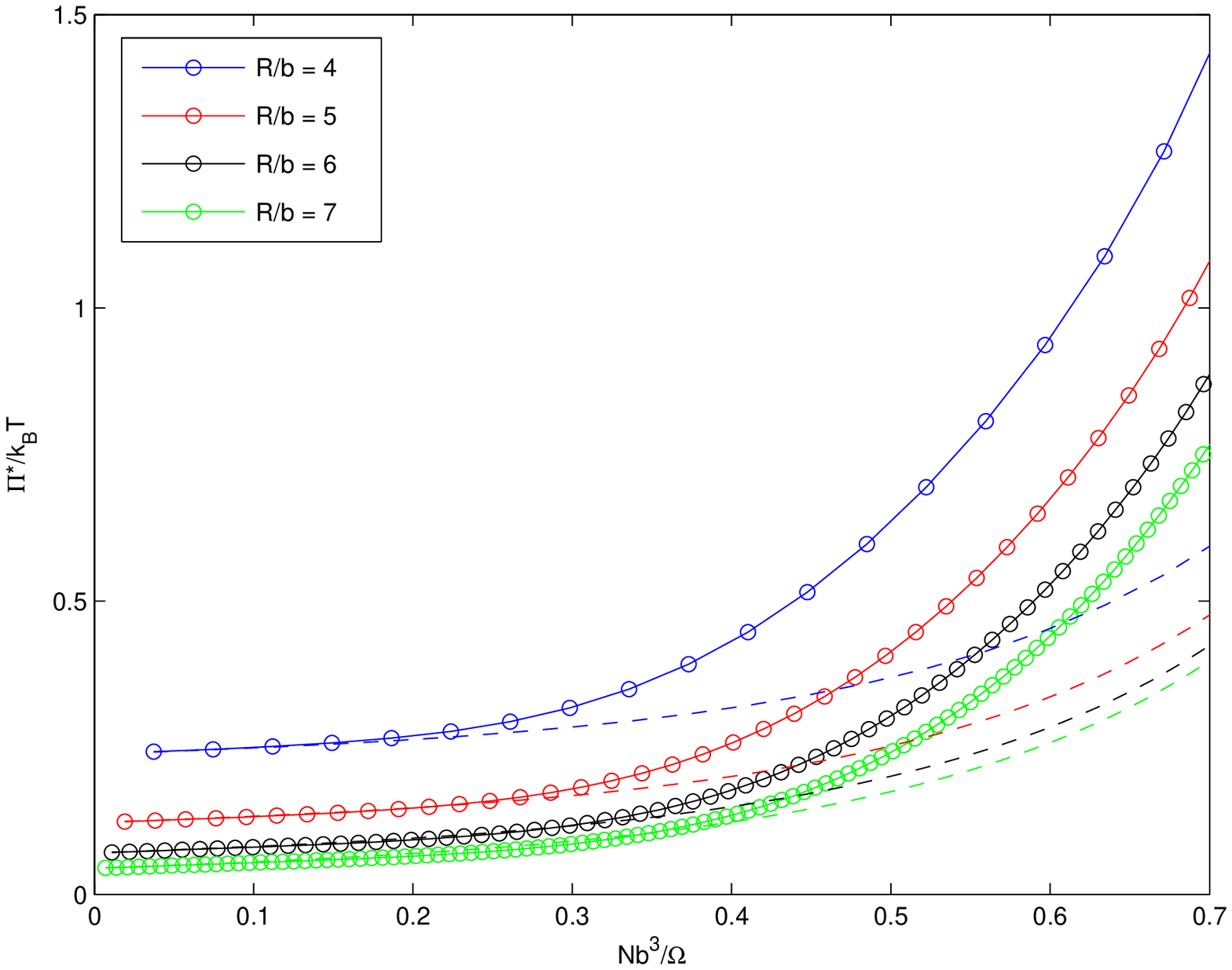}  
    \end{minipage}
\caption{}\label{osmotic}
\end{center}
\end{figure}

\begin{figure}[h]
\begin{center}
    \vspace*{1.0cm}
      \begin{minipage}[c]{15cm}
     \includegraphics[width=15cm]{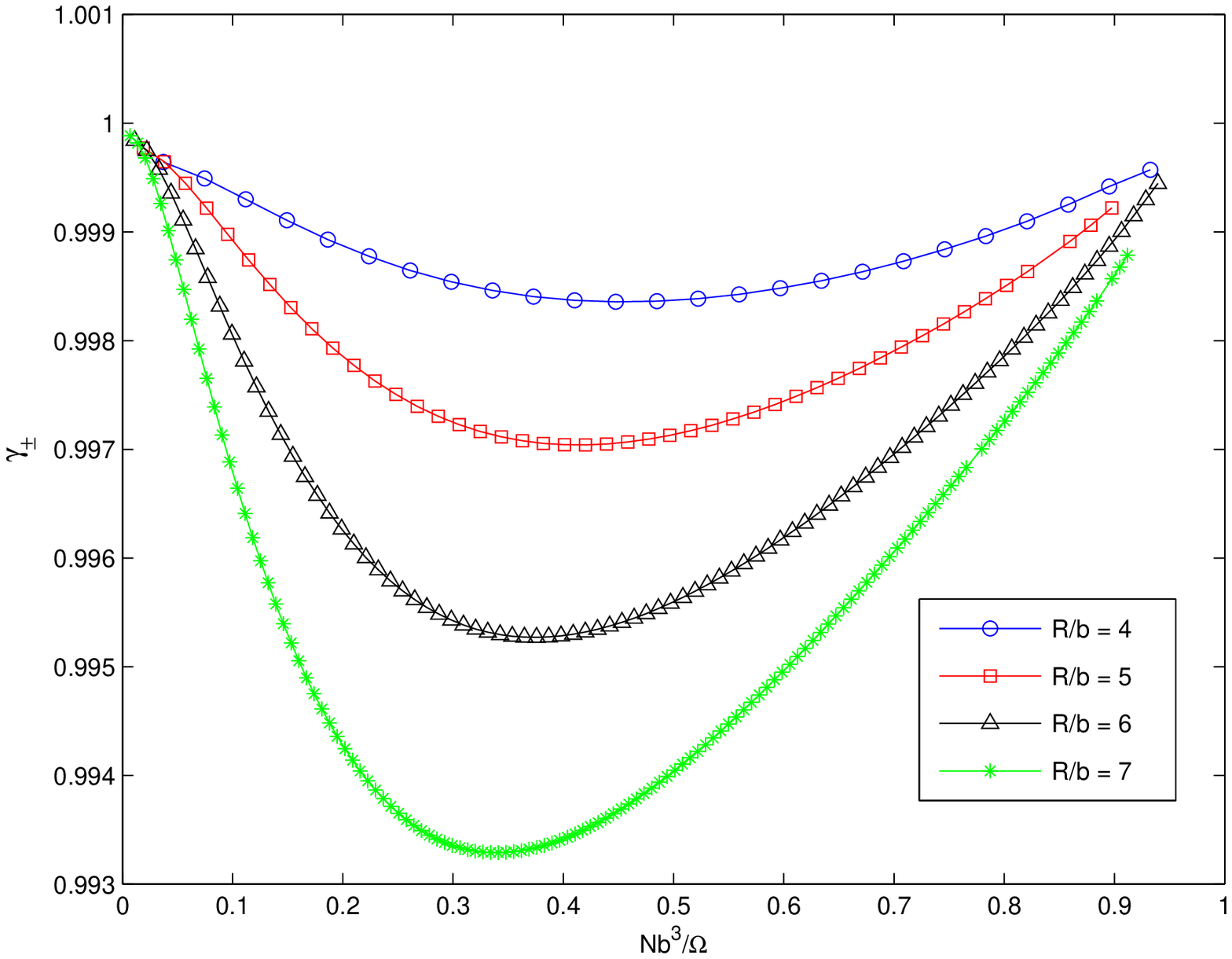}
    \end{minipage}
\caption{}\label{activity}
\end{center}
\end{figure}

\begin{description}
\item[Fig. 6.:] Difference in free energy of the
spherical cavity with and without polyelectrolyte chain, $\Delta F^{\star} = F^{*} - F\{\bar{\rho}_{p} = 0\}$.  All other parameters are the same as in Fig. ~\ref{free_vs_n2r3}.
\end{description}

\begin{description}
\item[Fig. 7.:] Non-extensive nature of free energy of confinement. Parameters are the same as in Fig. ~\ref{free_vs_n2r3}.
\end{description}

\begin{description}
\item[Fig. 8.:] Comparison between osmotic pressure for the confined polyelectrolyte chain and the homogeneous
phase (cf. Eq. (~\ref{eq:osmotic_homo})). Solid lines correspond to the inhomogeneous case and dashed lines represent the homogeneous system.
Parameters are the same as in Fig. ~\ref{free_vs_n2r3}.
\end{description}

\begin{description}
\item[Fig. 9.:] Mean activity coefficients for monovalent salt as a function of monomer density of the polyelectrolyte. For comparison purposes, parameters have been chosen to be the same as in Fig. ~\ref{free_vs_n2r3}.
\end{description}

\newpage
\vspace*{1.0cm}

\newpage
\vspace*{1.0cm}

\newpage
\vspace*{1.0cm}

\newpage

\vspace*{1.0cm}

\newpage
\vspace*{1.0cm}

\newpage
\vspace*{1.0cm}

\newpage
\vspace*{1.0cm}

\newpage
\vspace*{1.0cm}

\newpage
\vspace*{1.0cm}

\end{document}